\documentclass[a4paper, 10pt]{article}
\pdfoutput=1 

\usepackage[usenames, dvipsnames, svgnames, table]{xcolor} 
\usepackage{jcapmod}
\usepackage{verbatim}
\usepackage{etoolbox}
\usepackage[T1]{fontenc} 
\usepackage[latin1, utf8]{inputenc}
\usepackage[
    textwidth=0.71\paperwidth,
    textheight=0.8\paperheight,
]{geometry}
\usepackage{float}
\usepackage{graphicx}
\usepackage{hyperref}
\usepackage{amsmath}
\usepackage{amssymb}
\usepackage{siunitx}
\usepackage{microtype}
\usepackage[shortlabels]{enumitem}
\usepackage{cancel}
\usepackage{amsbsy}
\usepackage{color}
\usepackage[capitalise, nameinlink]{cleveref}
\usepackage{breakurl}
\usepackage[normalem]{ulem}
\usepackage{mathtools}
\usepackage{csquotes}
\usepackage{lmodern}
\usepackage{bm}
\usepackage{dsfont}
\usepackage{bbold}
\usepackage{empheq}
\usepackage{caption}
\usepackage{xparse}
\usepackage{subcaption}
\usepackage{nicefrac}
\usepackage{lineno}
\usepackage{ragged2e}

\hypersetup{
    colorlinks=true,
    linkcolor=blue,
    urlcolor=blue,
    citecolor=blue,
    filecolor=blue,
}

\makeatletter
\g@addto@macro\bfseries{\boldmath}
\makeatother

\pdfpageheight\paperheight
\pdfpagewidth\paperwidth

\DeclareSIUnit\parsec{pc}

\newcommand{\DKL}{D_{\mathrm{KL}}}

\newcommand{\ibigl}[1]{%
    \mathopen{\raisebox{0.25ex}{%
        \scalebox{1}[1.]{$\bigl#1$}%
    }}%
}
\newcommand{\ibigr}[1]{%
    \mathclose{\raisebox{0.25ex}{%
        \scalebox{1}[1.]{$\bigr#1$}%
    }}%
}

\renewcommand*{\vec}[1]{\mathbf{#1}}
\newcommand*{\unitvec}[1]{\vec{\hat{#1}}}

\newcommand{\vn}{\ensuremath{\vec{n}}}
\newcommand{\vtn}{\ensuremath{\vec{\tilde{n}}}}

\newcommand{\vk}{\ensuremath{\vec{k}}}
\newcommand{\vkp}{\ensuremath{\vec{k}^{\prime}}}
\newcommand{\vkn}{%
    \ensuremath{\vec{k}_{\vec{n}^{\vphantom{\prime}}}}%
}
\newcommand{\vknp}{%
    \ensuremath{\vec{k}_{\vec{n}^{\prime}}}%
}
\newcommand{\vktn}{%
    \ensuremath{\vec{k}_{\vec{\tilde{n}}^{\vphantom{\prime}}}}%
}

\newcommand{\kn}{\ensuremath{k_{\vec{n}}}}

\newcommand{\q}{\ensuremath{\vec{q}}}
\newcommand{\qp}{\ensuremath{\vec{q}^{\prime}}}

\newcommand{\qm}{%
    \ensuremath{\vec{q}_{\vec{m}^{\vphantom{\prime}}}}%
}
\newcommand{\qmp}{%
  \ensuremath{\vec{q}_{\vec{m}^{\prime}}}%
}

\newcommand{\vx}{\ensuremath{\vec{x}}}
\newcommand{\vxo}{\ensuremath{\vec{x}_{0}}}

\newcommand{\vr}{\ensuremath{\vec{r}}}

\newcommand{\vz}{\vec{0}}

\newcommand{\T}[2]{\vec{T}_{#1}^{#2}}

\newcommand*{\mat}[1]{\bm{\mathsf{#1}}}
\newcommand{\identity}{\ensuremath{\mathds{1}}}

\DeclareMathOperator{\sinc}{sinc}

\DeclareMathOperator{\Spec}{Spec}

\newcommand{\dd}{\mathrm{d}}
\newcommand{\ddc}{\dd^{3}}

\newcommand{\txo}{\tilde{x}_{0}}
\newcommand{\tyo}{\tilde{y}_{0}}

\newcommand{\nx}{n_{x}}
\newcommand{\ny}{n_{y}}
\newcommand{\nz}{n_{z}}

\newcommand{\reclat}{\mathrm{RL}}

\newcommand{\bigo}{\mathcal{O}}

\newcommand{\smallin}{%
    \mathrel{\scriptscriptstyle\in}%
}

\newcommand{\setN}{\mathcal{N}}

\newcommand*{\E}[1]{\texorpdfstring{\ensuremath{E_{#1}}}{E#1}}
\newcommand*{\Ehom}[1]{\texorpdfstring{\ensuremath{E_{#1}^{(1)}}}{E#1}}

\newcommand*{\Espace}{\texorpdfstring{\ensuremath{E^3}}{E(3)}}

\newcommand{\slabi}{\texorpdfstring{\ensuremath{\E{16}^{(\mathrm{i})}}}{E16i}} 

\newcommand*{\eigm}[2]{\Upsilon_{#1}^{#2}}

\newcommand{\Kdelta}{\delta^{(\textrm{K})}}

\newcommand{\Ddelta}{\delta^{(\textrm{D})}}

\newcommand{\LLSS}{L_{\mathrm{LSS}}}
\newcommand{\obs}{\mathrm{obs}}

\newcommand{\As}{A_\mathrm{s}}
\newcommand{\ns}{n_\mathrm{s}}

\newcommand{\dnc}{\ensuremath{d_{\textsc{nc}}}}
\newcommand{\dlss}{\ensuremath{d_{\textsc{lss}}}}

\newcommand{\Lobs}{L_{\obs}}
\newcommand{\fobs}{f_{\obs}}

\newcommand{\PR}{\mathcal{P}^{\mathcal{R}}}

\newcommand{\tphi}{\tilde{\phi}}
\newcommand{\tphiF}{\tilde{\phi}_{F}}

\newcommand{\tW}{\ensuremath{\widetilde{W}}}
\newcommand{\tWconj}{\ensuremath{\widetilde{W}}^{*}}

\newcommand{\nmax}{n_{\mathrm{max}}}

\newcommand{\kmax}{k_{\mathrm{max}}}
\newcommand{\ks}{k_{\star}}

\newcommand{\mmax}{m_{\mathrm{max}}}

\newcommand{\twod}{\mathrm{2d}}
\newcommand{\thrd}{\mathrm{3d}}

\newcommand{\gen}{\mathfrak{g}}
\NewDocumentCommand{\g}{ m g O{} }{%
  \ensuremath{%
    \IfValueTF{#2}
      {\IfBlankTF{#3}
         {\gen_{#1}^{#2}}
         {\left(\mkern-1mu\gen_{#1}^{#2}\mkern-1mu\right)^{\!#3}}}
      {\IfBlankTF{#3}
         {\gen_{#1}}
         {\left(\!\gen_{#1}\!\right)^{\!#3}}}%
  }%
}
\RenewDocumentCommand{\L}{ O{} m g }{%
  \IfValueTF{#3}
    {{L_{#2}^{#1}}_{#3}}
    {L_{#2}^{#1}}%
}
\newcommand{\M}[2]{\mat{M}_{#1}^{#2}}
\newcommand{\iM}{\mat{M}}
\newcommand{\iMT}{\mat{M}^{T}}
\newcommand*{\iMTp}[1]{%
    \ensuremath{%
        {\vphantom{\mat{M}}\smash[t]{\ibigl(\iMT\ibigr)}}^{\hspace*{-0.5pt}#1}%
    }%
}

\allowdisplaybreaks

\makeatletter
\DeclareRobustCommand{\rcite}[1]{%
  \rcite@aux#1,\@nil{#1}%
}
\def\rcite@aux#1,#2\@nil#3{%
  \if\relax#2\relax
    Ref.~\cite{#3}%
  \else
    Refs.~\cite{#3}%
  \fi
}
\makeatother

\let\orgautoref\autoref
\renewcommand{\autoref}{%
    \def\equationautorefname{Eq.}%
    \def\figureautorefname{Fig.}%
    \def\sectionautorefname{Section}%
    \def\subsectionautorefname{Section}%
    \def\subsubsectionautorefname{Section}%
    \orgautoref
}

\usepackage[table]{xcolor}

\definecolor{lightgray}{gray}{0.9}
\definecolor{Amber}{rgb}{1.0, 0.75, 0.0}
\definecolor{blizzardblue}{rgb}{0.67, 0.9, 0.93}
\definecolor{burningsand}{RGB}{220, 148, 129}
\definecolor{burgundy}{rgb}{0.5, 0.0, 0.13}
\definecolor{navy}{HTML}{003366}
\newcommand{\I}{\mathrm{i}}

\subheader{IFT-UAM/CSIC-26-115}

\title{Cosmic topology. Part Va. Information content of the observable Universe}

\collaboration{COMPACT Collaboration}
\collaborationImg{\includegraphics[width = 0.1\textwidth]{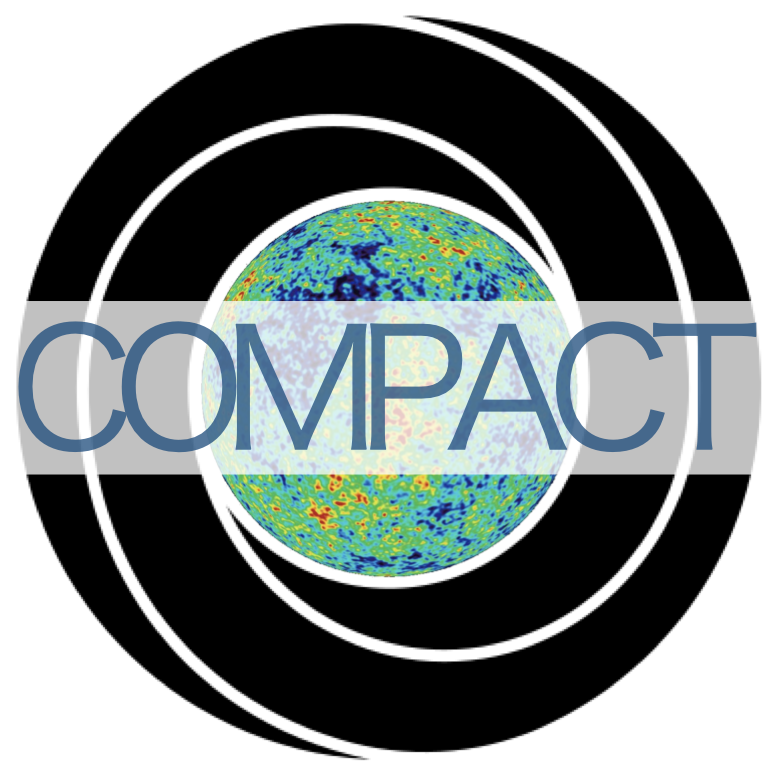}}

\author[a,b]{Deyan P. Mihaylov\textsuperscript{\(\dagger\)}}
\begingroup

\footnotetext[2]{Corresponding author.}
\endgroup
\author[c,d]{Mikel Martin Barandiaran,}
\author[e,f,g]{Stefano Anselmi,}
\author[c]{Javier Carron Duque,}
\author[a]{Glenn D. Starkman,}
\author[c,a,h]{Yashar Akrami,}
\author[a]{Craig~J.~Copi,}
\author[h]{Andrew H. Jaffe,}
\author[i]{Arthur Kosowsky,}
\author[j,a]{James B. Mertens,}
\author[a]{Anna Negro,}
\author[c]{George Alestas,}
\author[k]{Ricardo G. Rodrigues}
\author[k]{Thiago S. Pereira,}
\author[a]{Amirhossein Samandar,}
\author[l,a]{Andrius Tamosiunas,}
\author[a]{Cynthia Trendafilova}

\affiliation[a]{CERCA/ISO, Department of Physics, Case Western Reserve University, 10900 Euclid Avenue, Cleveland, OH 44106, USA}
\affiliation[b]{Department of Astronomy, Faculty of Physics, Sofia University ``St Kliment Ohridski'', 5 James Bourchier Blvd, 1164 Sofia, Bulgaria}
\affiliation[c]{Instituto de F\'isica Te\'orica (IFT) UAM-CSIC, C/ Nicol\'as Cabrera 13-15, Campus de Cantoblanco UAM, 28049 Madrid, Spain}
\affiliation[d]{Departamento de Física Teórica, Universidad Autónoma de Madrid, 28049 Madrid, Spain}
\affiliation[e]{INFN, Sezione di Padova, via Marzolo 8, I-35131 Padova, Italy}
\affiliation[f]{Dipartimento di Fisica e Astronomia ``G. Galilei'', Universit\`a degli Studi di Padova, via Marzolo 8, I-35131 Padova, Italy}
\affiliation[g]{Laboratoire Univers et Th\'eories, Observatoire de Paris, Universit\'e PSL, Universit\'e Paris Cit\'e, CNRS, F-92190 Meudon, France}
\affiliation[h]{Astrophysics Group \& Imperial Centre for Inference and Cosmology, Department of Physics, Imperial College London, Blackett Laboratory, Prince Consort Road, London SW7 2AZ, United Kingdom}
\affiliation[i]{Department of Physics and Astronomy, University of Pittsburgh, 3941 O'Hara Street, Pittsburgh, PA 15260, USA}
\affiliation[j]{David A. Dunlap Department of Astronomy \& Astrophysics, University of Toronto, 50 St. George Street, Toronto, ON M5S 3H4, Canada}
\affiliation[k]{Departamento de F\'{i}sica, Universidade Estadual de Londrina, Rod. Celso Garcia Cid, Km 380, 86057-970, Londrina, Paran\'{a}, Brazil}
\affiliation[l]{Institute of Theoretical Astrophysics, P.O. Box 1029 Blindern, N-0315 Oslo, Norway}

\emailAdd{info@compactcollaboration.org}
\emailAdd{deyan.mihaylov@case.edu}

\abstract{
If the Universe possesses a compact spatial topology with characteristic scale not much larger than the diameter of the observable Universe, then the associated symmetry transformations imprint specific statistical correlations on density perturbations. 
We derive the covariance between perturbation modes in the full topology volume and compute the induced covariance between modes in the observable subvolume.
We then use this covariance to estimate the available information about topology contained in the full set of observable linear perturbations. 
For two sample Euclidean examples---the cubic three-torus ($E_1$) and a three-torus with a quarter turn ($E_3$)---the three-dimensional density field carries enough information to detect topology on scales up to 25 percent 
larger than those accessible from the two-dimensional cosmic microwave background anisotropies alone. 
Future probes of the cosmic density field using deep galaxy surveys or neutral-hydrogen 21-cm intensity mapping, combined with computationally intensive searches over the large parameter space of admissible topologies, thus offer a plausible avenue to extending the range of detectable cosmic topologies.
}

\begin{document}
\maketitle
\noindent 

\flushbottom

\section{Introduction}
\label{sec:introduction}
What is the topology of the Universe and how can we use all the information available to us to find out? 
The most powerful searches to date have used the statistical properties of the temperature fluctuations in the cosmic microwave background (CMB).
However, this probe gets all its topological signal from the fluctuations in the geometry and stress-energy on the last scattering surface (LSS)---the spherical annulus centered on the observer from which CMB photons have free-streamed to us.
It largely fails to extract relevant information along the line of sight between the last scattering surface and us, and is therefore insensitive to the vast majority of modes that encode topological information.
The ongoing and anticipated galaxy surveys and line-intensity-mapping surveys have the potential to probe a much larger fraction of the volume of the universe than ever before, and will therefore be sensitive to many many more of the modes that are potentially interesting for discovering cosmic topology. 
How useful can we expect them to be?

The theory of general relativity dictates the \textit{local} geometry of spacetime but not its \textit{global} topology \cite{Ellis:1971, Lachieze-Rey:1995qrb, Levin:2001fg}.
Observations are consistent with a perturbed Friedmann-Lema\^{i}tre-Robertson-Walker (FLRW) metric, so the Universe is nearly homogeneous and isotropic on large scales. 
This constrains the background spatial geometry to constant curvature---Euclidean (\(E^{3}\)), spherical (\(S^{3}\)), or hyperbolic (\(H^{3}\))---but does not require that we live in the familiar simply connected covering space of any of them. 
Each of these geometries admits a rich variety of manifolds, obtained from that covering space by identifying points related by a discrete, freely acting subgroup of its isometry group \cite{Thurston:1982zz, wolf2011spaces}. 
For example, flat FLRW geometry admits 18 such classes, \E{1} through \E{18}, of which 17 are multiply connected;
each allows for a continuous family of manifolds parametrized by the lengths and directions of the translation vectors associated with the generating transformations of the group \cite{Lachieze-Rey:1995qrb, COMPACT:2023rkp, COMPACT:2025adc}. 
Which, if any, describes our Universe is an open question in observational cosmology \cite{Lachieze-Rey:1995qrb, COMPACT:2022gbl, COMPACT:2026uzi}.

There is little theoretical reason to exclude a multiply connected Universe. 
Quantum gravity generically permits topology-changing processes in the very early Universe \cite{HAWKING1978349, Dowker:2002hm, Gibbons:2011dh}, and some models favor a non-trivial topology outright \cite{Zeldovich:1984vk}. 
Others have argued that compact flat or hyperbolic spaces also have attractive features as initial conditions for inflation \cite{Cornish:1996st, 
Linde:2004nz}. 
The principle objection to searching for topology is that typical inflationary models predict too many e-foldings of growth for the effects of topology to remain visible.
The same objection would apply to curvature, but cosmologists are acculturated to searching for it.
Topology is best treated as a physical property to be constrained observationally, rather than fixed \textit{a priori} by an assumption of simple connectedness.

From an observational standpoint, several features of the cosmic microwave background (CMB) are at odds with the assumption of statistical isotropy \cite{Copi:2010na,
Planck:2019evm}.
Whether these anomalies share a common physical origin remains to be explained, but a non-trivial cosmic topology is among the best-motivated candidates for a genuine violation of statistical isotropy, since it alters only the global boundary conditions satisfied by the fields and leaves the local equations of motion untouched \cite{Levin:2001fg, Riazuelo:2003ud, Fabre:2013wia, COMPACT:2023rkp, COMPACT:2024cud, CarronDuque:2026a}.

Concretely, a non-trivial topology imposes periodic identifications on space and all its contents.  This can be visualized as tiling the covering space with a fundamental domain---for example a parallelepiped, in the flat case.
Equivalently, and more usefully, it is a set of discrete symmetries that every field must satisfy,\footnote{
    The fundamental domain is a convenient visualization but carries no physical reality: any domain that tiles the covering space under the same set of symmetry transformations is observationally equivalent.
} including the cosmological perturbations about the homogeneous and isotropic background and all forms of stress-energy. 
These perturbations are conventionally expanded in a Fourier basis, not merely for convenience but because the linearized equations of motion evolve each Fourier mode independently. 
In place of the continuum of Fourier modes available in the $E^3$ covering space, the eigenmodes of the Laplacian on a compact flat manifold are discrete linear combinations of Fourier modes selected by the symmetries of the topology. 
Because those symmetries are themselves anisotropic, so is the resulting discretization, e.g. the allowed wavelengths along a direction pointing toward a face of the fundamental domain differ from those along a direction pointing toward a corner. 
This anisotropy is imprinted on cosmological observables, which cease to be statistically isotropic and, for most topologies, statistically homogeneous or parity preserving as well; its degree and pattern depend on the topology, its scale, and the position and orientation of the observer within the space \cite{Riazuelo:2003ud, Niarchou:2007nn, COMPACT:2023rkp, COMPACT:2025adc}. 

Since the Wilkinson Microwave Anisotropy Probe (WMAP), the primary observational probe of cosmic topology has been the pattern of anisotropies in the CMB. 
Its advantage over the leading alternative---searching for ``topological clones,'' that is, instances in which the same astrophysical system (e.g., a group of quasars) is observed in more than one direction because of the multi-connectedness of space \cite{Fagundes:1987, Lehoucq1996, Uzan:1999de}---is that such clones are seen at different ages and from distinct viewing angles, making the identifications challenging. 
Related techniques such as cosmic crystallography are sensitive to selection effects and uncertainty in redshift measurements. 
The primary CMB,\footnote{
    As opposed to secondary anisotropies, e.g.\ Sunyaev-Zel'dovich scattering or lensing effects, which do probe the interior.
} by contrast, consists of photons that last scattered within a fractionally narrow range of redshifts. 
Its temperature and polarization fluctuations are therefore sourced by well-understood linear perturbations of the geometry and stress-energy on a thin spherical shell centered on the observer, namely the last scattering surface (LSS), of diameter \(\dlss \simeq \qty{28}{\giga\parsec}\), several times the Hubble diameter \(2 c / H_{0} \simeq \qty{8.6}{\giga\parsec}\). 
The relation between the statistics of the underlying cosmological fields, which carry the topological information, and the statistics of CMB observables is straightforward, and the utility of the CMB as probe of topology has grown as map resolution and signal-to-noise have improved.

Two distinct CMB strategies have been pursued. 
One of them searches for matching pairs of circles in the sky, along which the same physical points on the LSS are observed in different directions \cite{Cornish:1996kv, Cornish:1997hz, Cornish:2003db}; no statistically significant matches were found in either WMAP or \textit{Planck} \cite{Cornish:2003db, ShapiroKey:2006hm, Vaudrevange:2012da, Planck:2015gmu}, implying that any closed geodesic through our position exceeds \(\sim 98.5\%\) of \(\dlss\). 
The second approach examines the covariance matrix of the harmonic coefficients \(a_{\ell m}\), which retains the correlations discarded by the rotationally averaged angular power spectrum \cite{Riazuelo:2003ud, Riazuelo:2002ct, Phillips:2004nc, Kunz:2006, Niarchou:2007nn, Fabre:2013wia}. 
These correlations persist past the regime in which circles are visible, and remain measurable in principle even when the topology scale exceeds \(\dlss\) \citep{Fabre:2013wia, COMPACT:2023rkp}.

Nevertheless, anisotropies realized on a two-dimensional surface can encode only a fraction of the information available in principle. 
Suppose that the topological information resides primarily in modes with wavevectors \(\vk\) of magnitude below some \(\kmax\). 
The LSS then supports \(N_{\twod} = \bigo\left(\smash{\left(\kmax \dlss\right)^{2}}\right)\) independent modes, while its interior---the full volume accessible to an observer---supports \(N_{\thrd} = \bigo\left(\smash{\left(\kmax \dlss\right)^{3}}\right)\). 
If the shortest closed loop through our position is longer than \(\dlss\), previous work finds the relevant information concentrates in multipoles \(\ell \lesssim 30\) \cite{COMPACT:2022gbl, COMPACT:2023rkp}, so \(N_{\twod} \simeq 900\) and \(N_{\thrd} / N_{\twod} \simeq 30\). 
For a shorter loop, informative CMB modes extend to much higher \(\ell\), and the three-dimensional advantage is larger still. 
The physical origin of the loss is projection: many three-dimensional modes contribute to the same angular mode on the LSS, erasing part of the anisotropy pattern that topology imprints.

The relevance of this shortfall is highlighted by the fact that the three-dimensional density field is about to be mapped over a large fraction of the observable Universe by large-scale-structure surveys:\footnote{
    By large-scale-structure surveys we mean any observations of tracers of the matter density or velocity fields---galaxy surveys, peculiar-velocity surveys, line-intensity mapping, etc.
} spectroscopic and photometric galaxy surveys including DESI, {\it Euclid}, SPHEREx, and Rubin-LSST \cite{DESI:2025fxa, Euclid:2024yrr, Bock:2025ijl, LSSTScience:2009jmu}, and by line-intensity-mapping surveys including CHIME, CHORD, HIRAX, SKAO, COMAP, and proposed Stage-II experiments \cite{CHIME:2025cee, Vanderlinde:2019tjt, Newburgh:2016mwi, Majumdar:2026dgt, Stutzer:2024rps, CosmicVisions21cm:2018rfq}. 
Reaching \(z \sim 27\), as 21-cm surveys may \cite{EorCDScienceWorkingGroup:2026trp}, corresponds to \(\sim 82\%\) of the comoving distance to the LSS. 
The trade-off against the CMB is between a three-dimensional volume at lower redshift and a two-dimensional surface at the highest accessible one -- which probe places a more stringent constraint on topology depends on the survey.

In this work we demonstrate that an ideal (noise-free and complete) three-dimensional survey does contain substantially more information about cosmic topology than the CMB. 
We build upon the formalism developed in the COMPACT collaboration for the CMB, which characterized the Laplacian eigenmodes and covariance matrices of Fourier modes for the 18 Euclidean topologies \cite{COMPACT:2023rkp, COMPACT:2024cud, COMPACT:2025adc}.
We derive the covariance between perturbation modes induced by non-trivial topology in the full topology volume, and compute the correlations this induces between pairs of modes within the smaller observed volume, for the cubic three-torus (\E{1}) and the three-torus with a quarter turn (\E{3}). 
We quantify how non-trivial topology alters the statistical properties of the matter density field, and assess the detectability of these signatures as a function of the parameters of the topology, the position and orientation of the observer, and the volume observed. 
This study provides the first systematic treatment of topological signatures in the three-dimensional density field, and it establishes that topology is detectable on scales substantially larger than the CMB alone permits.

In \autoref{sec:information}, we discuss the information content of the observable Universe, and  present the advantages of what we define as a “Complete Primordial Survey” for assessing the ability of the three-dimensional density field to probe cosmic topology. 
In \autoref{sec:theory}, we introduce the theoretical concepts and the formalism, including the signatures of topology in the statistical properties of the matter density field. 
In \autoref{sec:methodology} we explain the observer window function, how it affects the measured modes, and the covariance of these modes under different topologies. 
In \autoref{sec:results} we discuss the numerical results for select topological parameters, computing the total amount of information about topology in the measured modes, and highlighting general trends and behaviors, as well as comparing the large scale structure with the CMB as a probe for cosmic topology. 
In \autoref{sec:conclusions} we summarize our main findings and discuss future directions.

\section{Information content of the observable Universe}\label{sec:information}

The goal of this investigation is to quantify the extent to which cosmological observations can distinguish a universe with non-trivial topology from its simply-connected covering space. To this end, given a set of observables and two competing models (i.e.~two different topologies), each with an associated probability distribution for the data, we employ the Kullback--Leibler (KL) divergence to quantify the distinguishability between them (more details in \Cref{ssec:kl}).

In the following, we argue that, under specific working assumptions, it is possible to estimate an approximate upper bound on the detectability we can obtain from all possible observations of cosmological fluctuations (e.g.~galaxy number counts and CMB temperature fluctuations). This is crucial because weak discriminating power would constitute a no-go result for searches for topology based on cosmological perturbations. In particular, we present a step-by-step argument identifying a convenient space and field for which the KL divergence between a non-trivial topology and its covering space (or vice versa) is the largest attainable within our past light cone.

\subsection{Cosmological and statistical framework}

We work within a standard cosmological framework, in which primordial fluctuations in the cosmologically relevant fields can conveniently be expressed as superpositions of eigenmodes of the Laplacian operator.
We will focus on scalar fluctuations because they have been observed, though primordial tensor fluctuations, should they ever be observed, potentially also contain sizable amounts of information \cite{COMPACT:2024cud,Samandar:2025kuf}. 
We will also, for the purposes of this paper, consider only Euclidean manifolds.

Within this framework, we assume that the amplitudes of linearly independent eigenmodes are statistically independent Gaussian random variables, as predicted by standard inflationary cosmology. 
Moreover, imposing Bunch--Davies initial conditions selects the temporal mode function associated with each independent spatial eigenmode, so that the scalar curvature perturbation field contains all the required initial information and no independent specification of the corresponding velocity perturbations is required.
The amplitude of each late-time cosmological mode is fully determined,  at linear order, by the corresponding initial adiabatic curvature-perturbation mode amplitude and the appropriate transfer function. Following standard practice and without claiming that this assumption follows from inflation, we adopt the standard statistical characterization of the Laplacian-eigenmode amplitudes and assume that the scalar-fluctuation power spectrum is nearly scale-free.

Standard inflationary cosmology is conventionally formulated using covering-space boundary conditions. 
We will adopt this same framework for the statistics of primordial fluctuations, even though  the conventional inflationary paradigm is altered by observable topology, precisely as it would be by observable spatial curvature.
Specifically, observation of topology would imply that physics beyond standard inflation is required.
Nevertheless, one might imagine that inflation takes place on a pre-homogenized manifold and is responsible, in the usual way, for the generation of primordial fluctuations. 
Again, as in a curved manifold, 
topology-induced quantum effects could also alter the early-Universe dynamics and thereby modify the primordial power spectrum \cite{COMPACT:2026vdj}. 
Under these working assumptions, the topological signature is encoded in the Laplacian eigenmodes permitted by the isometries of the specific topology.
For flat (Euclidean) manifolds, the eigenmodes of the scalar Laplacian are a discrete subset of the usual Fourier modes of the covering-space, or a finite linear combination of such Fourier modes with the same wave number.
Since this is the only effect of topology, the transfer functions are functionally identical to those in the covering space, except that they are now sampled at discrete wavenumbers.

Each  eigenmode evolves independently (to leading order in perturbation theory), so transfer functions act as linear transformations in Fourier space.\footnote{
    Real-space observables are obtained by convolution with a kernel, which is also a linear operation.} Of course, for some cosmological observables, parameter values, and ranges of scales, linear perturbation theory is not sufficiently accurate. 
We avoid that regime, and, fortunately, find that a great deal of information is contained in scalar fluctuations that remain in the linear regime to this day.

\begin{figure}[t]
    \centering
    \includegraphics[scale=1]{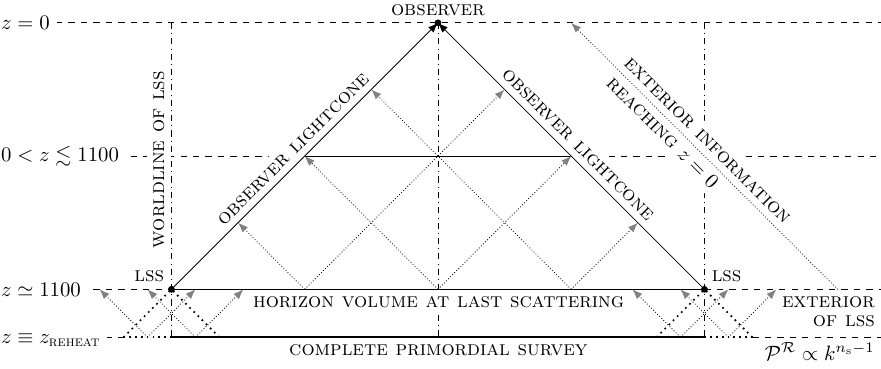}
    \caption{%
        Pictorial representation of the information flow from an initial spatial hypersurface to the interior of our lightcone. 
        The observer at $z = 0$ collects information from their past light cone 
        (and potentially its interior) 
        extending back to the maximum accessible redshift $z_{\textsc{LSS}}\simeq 1100$, the  last scattering surface (LSS). 
        By causality and unitarity, any information on that past lightcone was also encoded in the values of the fields describing observables (and the time derivatives of those fields) on the horizon volume at last scattering, i.e. the closed observer-centered ball $z\leq z_{\textsc{LSS}}$ bounded by the LSS.
        No information  exterior to the lightcone reaches the observer.
        In principle, the observer could look even further back  to $z_{\textsc{reheat}}$, the epoch of reheating at the end of inflation, but the additional volume (and information) is very small. 
        A complete observational survey out to $z_{\textsc{LSS}}$ is therefore equivalent to a complete primordial survey on the observer-centered ball $z\leq z_{\textsc{reheat}}$. 
    The thick solid line at $z_{\textsc{reheat}}$ represents a spatial cross-section of the complete primordial survey.%
    }\label{fig:lightcone}
\end{figure}

\subsection{Discriminating power of three-dimensional surveys}
In this setting, fluctuations of some observed cosmological field (e.g.~the CMB temperature field) are key observables for discriminating between different topologies. 
In a realistic observational framework, combining a larger number of cosmological probes  increases the discriminating power we can reach.
We therefore seek, ultimately, to combine the broadest possible range of cosmological observations to maximize the prospects of detecting a possible non-trivial topology of the Universe. 
In future work, we will provide assessments of the topological information accessible using existing, ongoing, planned, or realistically foreseeable cosmological surveys.
However, if our goal is to estimate an upper bound on the discriminating power we could reach, we can simplify the computation. 
We consider ideal experimental conditions, with no noise, no observational systematics (e.g., foregrounds), no sky mask, and the ability to exactly model other relevant effects. Within the standard cosmological paradigm, if the cosmological parameters were known exactly, nonlinearities were negligible, and the full three-dimensional (3D) matter-density field\footnote{
    Velocity information is redundant for Bunch-Davies initial conditions.} on our past light cone were accessible in angular and redshift coordinates, this field would provide the maximum possible discriminating power between a model with non-trivial topology and the corresponding covering-space model, 
at least until a comparable survey of the tensor field could be completed.

Under these assumptions, because the fluctuations are adiabatic, other species (e.g.~radiation) provide no additional information about topology. As long as linear perturbation theory remains valid, the evolution of perturbations of every species is related to the primordial curvature perturbations through linear, invertible mappings (i.e.~transfer functions), which neither remove nor add information when the cosmological parameters are known exactly. Consequently, any energy-density field for which the full 3D map is known carries the same information as the matter field. By contrast, projecting this field onto the sky will contain only a subset of that information.

Therefore, the KL divergence of the 3D matter field on our past light-cone, between two models having the same values for the background cosmological parameters but different boundary conditions (one corresponding to a non-trivial topology and the other to its covering space) represents the maximum discriminating power we can reach using cosmological perturbations. Consistently with our objective of deriving an upper bound on our ability to detect non-trivial topology, we also fix the topological parameters to specified values. 

\subsection{Quantifying topological information: the complete primordial survey}

Even under all the preceding assumptions, a direct computation of the KL divergence would be cumbersome. 
A convenient property of the KL divergence, however, allows the same result to be obtained much more simply: it is invariant under bijective maps (e.g., \cite{kullback1997information} Ch. 2.4). 
Provided that nonlinearities can be neglected, the mapping from the comoving curvature perturbation \(\mathcal{R}\) on appropriate finite-volume 3D spatial hypersurfaces--defined by the background metric and cosmic time--to the matter perturbations on our past light cone is linear and invertible. 
Therefore, as illustrated in Figure~\ref{fig:lightcone} (and described in the figure caption), we may instead compute the KL divergence of the curvature perturbations within a region of a 3D spatial hypersurface at the epoch of last scattering, with a comoving volume equal to that enclosed by the last-scattering surface. 
In practice, if we consider a time shortly after reheating, the corresponding difference in comoving volume is negligible, and we can simply use the standard dimensionless power spectrum of the comoving curvature perturbation $\mathcal{R}$,\footnote{
    As shown in \cite{Bardeen:1980kt,Lyth:1984gv,Wands:2000dp}, if during inflation cosmological fluctuations are adiabatic, \(\mathcal{R}\) is conserved outside the horizon during the reheating era.}
conventionally written as
    \begin{equation}
    \PR(k) = \As \left(\frac{k}{\ks}\right)^{\!\ns-1} ,
    \end{equation}
where the scalar amplitude $\As$ is defined at the fiducial wavenumber $\ks$ and $\ns$ is the scalar spectral tilt. 
We refer to this region of the 3D spatial hypersurface at the epoch of reheating/last-scattering as the Complete Primordial Survey (CPS). 
The spatial cross-section of the CPS shown in Figure~\ref{fig:lightcone} is indicated by the thick solid line.

Spatial topology admits identifications that preserve the local homogeneity and isotropy of the metric. 
These identifications are defined on the 3D spatial hypersurfaces determined by the FLRW metric and cosmic time. Moreover, the curvature perturbation and its power spectrum are gauge invariant on these hypersurfaces. Computing the KL divergence of $\mathcal{R}$ measured on the CPS in a topologically-non-trivial manifold compared to in the covering space is therefore a much simpler way of determining the topological discriminating power of the 3D matter field on our past light cone than trying to do the calculation directly on the past light cone.

To complete the argument, we must consider the sensitivity of the KL divergence calculation to long-wavelength modes. Within the standard cosmological framework, ${\mathcal{P}}^{\mathcal{R}}(k) \propto k^{\ns - 1} \simeq k^{0}$, which might appear to imply that the KL divergence is infrared divergent and hence unphysical. 
However,  these long wavelength modes represent fluctuations relative to the mean over spatial hypersurfaces of infinite extent, and in particular far beyond the reach of our past light cone, which real surveys inhabit, or the CPS, the compact portion of a spatial hypersurface interior to our past light cone which we consider.
As discussed below, on the CPS, as on the past light cone, we can readily avoid this non-physical infrared divergence.

By causality of the Einstein--matter equations on a weakly perturbed FLRW background, observable combinations of fields on or within the observer's past light cone can depend only on initial data specified on the CPS, the portion of the primordial hypersurface contained in their causal past. This region contains the maximum primordial information accessible to a given observer.

Within the CPS, we decompose $\mathcal R$ into its average $\overline{\mathcal R}_{\rm CPS}$ and fluctuations relative to that average. Even for a Euclidean spatial manifold with non-trivial topology, $\overline{\mathcal R}_{\rm CPS}$ need not vanish when the CPS does not contain the entire spatial volume, as in the case considered here. Moreover, the CPS homogeneous mode should not be identified with the globally homogeneous $k=0$ mode: for a finite CPS, it is in general a window-weighted combination of global Laplacian eigenmodes and may therefore contain some topology-dependent information. 
Nevertheless, from observations confined to the CPS, this homogeneous mode is, at leading order, observationally degenerate with a redefinition of the locally inferred homogeneous and isotropic background.\footnote{ 
    The average 3-Riemann tensor will not, in general, correspond to an isotropic 3-geometry, but will instead be given by one of the five anisotropic Thurston geometries.
    However, the deviations from isotropy are first order in the perturbations around the homogeneous isotropic  metric on the manifold, therefore their contributions to the mode equations for the fluctuations ${\mathcal R}$  within the CPS are 2nd order in perturbation theory. 
    Since we are studying only those modes that remain in the linear regime, we can self-consistently neglect the coupling of the fluctuations to the homogeneous mode and take the background metric to be Euclidean FLRW.}

We find that we are led, anyway, to neglect the homogeneous mode on the CPS when considering the information available to distinguish between a topologically non-trivial manifold and the covering space of its geometry.
This is because, in order to calculate the expected contributions to the homogeneous mode from fluctuations beyond the CPS, we need an infra-red cutoff to the power spectrum.
In the compact manifold, there is automatically a cutoff, given by the finite extent of the manifold; however, in the covering space, the cutoff depends sensitively on the origin of the fluctuations, e.g. the precise model of inflation.
Not wanting to rely on such specific models, we choose instead to remove the homogeneous mode and consider only the information in the fluctuations.

As discussed above, we already neglect topology-dependent modifications of the homogeneous early-Universe dynamics, including possible topology-induced quantum effects on the expansion history and primordial power spectrum, and adopt the usual nearly scale-free statistical characterization of the Laplacian-eigenmode amplitudes. 
Projecting out the locally background-like CPS homogeneous mode implements the same restriction at the level of the observable region, leaving us to quantify the topology-dependent information encoded in correlations of fluctuations about the flat-FLRW-approximation to the CPS mean.

Finally, in this paper we will replace the three-ball of the CPS by a cubic box at the same redshift, aligned with the topology but smaller in size. It is clear that there is no 1-to-1 map of the fields (and information) from this box to our past light cone. 
However, the topological information content of this box, if its size is appropriately chosen, should be a good estimate of the information on the CPS, and therefore on our past light cone. The main advantage of this choice is computational in two key aspects. First, the effect of the topology simplifies significantly, as the three-dimensional window kernels become separable into three one-dimensional ones whose symmetries can be exploited (\cref{ssec:obswin}). Secondly, one can ensure that no information from outside the CPS is present in the analysis by just dropping the zero mode (\cref{app:integralconstraint}).

The question remains what size cube to take to simulate the ball of the CPS.
A cube of side-length equal to the diameter of the CPS clearly provides an upper limit on the information content of the CPS.
We will identify an ``information-matched benchmark'' cube that provides the same information as the surface of the CPS for a manifold that is just large enough to fully contain the LSS;
this benchmark cube provides a reliable lower limit on the total information content of the CPS.

Quantifying the residual difference between the spatial and light-cone descriptions requires realistic mock catalogues including all observational effects, and lies beyond the scope of the present work. 

\section{Topology and three-dimensional observables}
\label{sec:theory}

In this section we first recall essential formalism for topology of Euclidean three manifolds, focusing on compact manifolds, since the semi-compact Euclidean manifolds can be regarded as limiting cases (with the exception of the rotated slab \slabi, which we will treat separately in future work).
In particular we discuss the eigenmodes of the Laplacian on compact Euclidean manifolds, their relation to Fourier modes, the standard eigenmodes on the covering space \Espace~ (or \E{18}),  and how that relation can be represented as a correlation matrix of Fourier mode amplitudes.
We next connect (in \cref{eqn:cqq_general}) that theoretical correlation matrix to the correlation between amplitudes of modes observable in a finite volume, representing the accessible three-volume of the observable universe, which we take for this work to be a cubic box, 
for the reasons discussed in \autoref{sec:information}.
Finally we provide specific examples of the observer-box mode-mode correlation matrix in three representative  three-manifolds: the simple three-torus \E{1} (\cref{eqn:Cqq_E1}), the quarter-turn space \E{3} (\cref{eqn:Cqq_E3}), and the covering space \E{18} (\cref{eqn:CE18}).

\subsection{Topology fundamentals}\label{ssec:top_fundamentals}

Any spatially flat three-manifold is constructed as a quotient of the usual infinite Euclidean space $E^3$.
However, not every quotient of the covering space yields a smooth manifold and one can show that only quotients induced by the action of discrete, freely acting and properly discontinuous subgroups of the $E^3$ isometry group will lead to smooth, flat quotient manifolds. 
In simpler terms, every flat three-manifold arises from identifying points in $E^3$ that are related by the action of some translations, corkscrew motions, or glide reflections.
This identification can be pictured as a set of symmetry conditions on all functions on the covering space, as a tiling of the covering space by suitably shaped and oriented domains, or as boundary conditions on such domains. 

The classification of all such Euclidean three-manifolds is well established.
There are 18 topologically non-equivalent classes of manifolds, typically labeled \E{1} through \E{18}.
For instance, \E{1} is the simple three-torus, \E{3} is the so-called \textit{quarter turn} space, and \E{18} is the infinite Euclidean covering space \(E^{3}\).
Within this classification, \E{1}--\E{10} are fully compact, with \E{1}-\E{6} being orientable manifolds and \E{7}-\E{10} non-orientable. 

The manifolds having each of these 10 compact topologies can be characterized by three linearly independent translation vectors, and thus by no more than nine real free parameters, of which three are equivalent to the orientation of the observer's coordinate system, leaving six.
Each translation vector \(\T{A_{j}}{\E{i}}\) is associated with one of the generators \(\g{A_{j}}{\E{i}}\) of the discrete subgroup \(\Gamma^{\E{i}}\) of the isometry group of \Espace ~(i.e., \(O(3) \times R^{3}\), rotations and translation in 3D) that characterizes the topology of \E{i},
\begin{equation}\label{eqn:Eigenerator}
    \g{A_{j}}{\E{i}}: \quad \vx \to \vxo + \M{A}{\E{i}}(\vx - \vxo) + \T{A_{j}}{\E{i}} \,,
\end{equation}
with \(\M{A}{\E{i}}\) an element of \(O(3)\) and \(\vxo\) a vector describing the location of the axis around which \(\M{A}{\E{i}}\) rotates or of the plane across which \(\M{A}{\E{i}}\) reflects.\footnote{
    Here \(A\) labels distinct \(\M{A}{\E{i}}\), while \(j\) labels distinct \(\T{}{\E{i}}\) associated with the same \(\M{A}{\E{i}}\).} 
Each \E{i} is associated with specific choices of \(\M{A}{\E{i}}\). 
(As discussed in \cite{COMPACT:2023rkp} and \cite{COMPACT:2025adc}, these choices are not unique, but different choices do not lead to different manifolds.)

We see from \cref{eqn:Eigenerator} that, in addition to the six real parameters describing the vectors $\T{A_{j}}{\E{i}}$, and the three Euler angles describing the orientation of the observer, there are up to three real parameters determining $\vxo$ -- or equivalently, the location of the observer.
In fact, the number of actual parameters characterizing a specific compact \E{i} manifold is less than six in most of the topologies because the three vectors cannot all be freely chosen except in \E{1}, and typically not all components of \(\vxo\) are physically material.

In addition to the 10 Euclidean topologies \E{1}--\E{10} whose manifolds are compact, there are \E{11}--\E{15}, which are infinite in one dimension, and \E{17}, which is infinite in two, but each of these can be regarded for cosmological purposes as a limit  of \E{1}--\E{10} (i.e., as the length of one or more of the translation vectors goes to infinity). 
Only \E{16} need be separately considered, because it is a slab where the identification is associated with an arbitrary rotation. (Only rotations by  multiples of \(\pi / 3\) and \(\pi / 2\) can be obtained as limits of one of the compact manifolds.) 
In the balance of this paper, we confine our attention to the compact Euclidean manifolds.

A key physical consequence of a non-trivial spatial topology is that it modifies the set of allowed eigenmodes of the scalar (and tensor) Laplacians. 
These modes determine how cosmological perturbations can exist within the manifold, thereby imprinting characteristic patterns on observable quantities such as the CMB or large-scale matter distribution.
A standard feature of the conventional inflationary cosmology is that the fluctuations in the fields sourcing  perturbations in stress-energy and the metric are Gaussian random fields. 
These fluctuations can be expressed as linear combinations of eigenmodes of the Laplacian with amplitudes that are Gaussian random statistically independent variables of zero mean. 

The usual basis of eigenmodes of the scalar Laplacian in \Espace~ is Fourier modes characterized by their wavevector $\vk$.
Non-trivial topology has two important effects on these eigenmodes.  
First it restricts the set of allowed \(\vk\) vectors -- in the compact manifolds, it replaces the 3D continuum of $\vk$ by the reciprocal lattice of the pure translations in $\Gamma^{\E{i}}$.
These pure translations generate an ``associated \E{1}'' (which we label \Ehom{i}) for each compact manifold, and thus the allowed wavevectors are in the reciprocal lattice $\reclat(\Ehom{i})$ of that associated \E{1} manifold.

Second, or all compact manifolds except \E{1}, the topology also enforces perfect correlations between amplitudes of Fourier modes whose wavevectors are related by the point group $P^{\E{i}}$ associated with the isometries in $\Gamma^{\E{i}}$ -- i.e. the discrete subgroup of $O(3)$ generated by the  $M^{\E{i}}_{A}$.

As described in \cite{COMPACT:2023rkp}, the scalar eigenmodes on \E{i} are thus of the form 
\begin{equation}
    \label{eqn:generaleigenmodeformula}
    \eigm{\vk}{\E{i}} (\vx) = \frac{1}{\big| P^{\E{i}}\big|^{1/2}} \sum_{\g{A_{j}}{\E{i}}} \exp \left(\I \vec{k}^T \g{A_{j}}{\E{i}} \vx\right) \,,
\end{equation}
with eigenvalue \(-\vk^{2}\). 
Here the sum is over a small, finite set of isometries \(\g{A_{j}}{\E{i}}\), one for each element of the point group $P^{\E{i}}$, while \(\big| P^{\E{i}} \big|\) is the order (i.e., number of elements) of this group. 
Eigenmodes can be labeled by any one of the wavevectors $(\M{A}{\E{i}})^T\vk$ appearing in the sum; we choose one and call it $\vk$.
The complete set of distinct eigenmodes are labeled by $\{\vkn / \vn\in \setN^{\E{i}}\}$, with $\vn$ a triplet of integers in an index set 
$\setN^{\E{i}}$ appropriate for each \E{i}. These are a subset of the full reciprocal lattice of the associated \E{1}. For a complete description of each Euclidean topology class and their harmonic properties, see \cite{COMPACT:2023rkp} for the orientable cases and \cite{COMPACT:2025adc} for the non-orientable cases.

Let $\phi(\vx)$ be a real scalar field defined on our spatial manifold and representing any of the usual cosmologically relevant fields, e.g. the matter density field or the primordial curvature perturbation field.
Then $\phi$ can be expanded in the \(\eigm{\vkn}{\E{i}}\) basis
\begin{equation}
    \phi(\vx) = \frac{1}{V_{\E{i}}} \sum_{\vn \in \setN^{\E{i}}} \tphi(\vkn) \eigm{\vkn}{\E{i}} (\vx)\,.
\end{equation}
However, in order to compare this to what we would measure if we lived in the covering space, we want to expand $\Upsilon^{\E{i}}_{\vkn}$ in Fourier modes as in \cref{eqn:generaleigenmodeformula}, and write
\begin{equation}
    \phi(\vx) = \frac{1}{V_{\E{i}}}\sum_{\vn \in \mathbb{Z}^3\backslash\{\vec{0}\}} \tphiF (\vkn) e^{\I \vkn^{T} \vx} \,.
\end{equation}
In the covering space the sum above would be replaced by an integral (and $V_{\E{i}}$ by $(2\pi)^{3}$), and,
for primordial fluctuations, the $\tphi_F$ would be independent Gaussian (or nearly Gaussian) random variables of zero mean.
In contrast,
in the compact topologies, \(\vkn\) live on $\reclat(\Ehom{i})$, and,
in all  except \E{1}, the $\tphiF (\vkn)$ of \(\vkn\) that are related by elements of $P^{\E{i}}$ are correlated but still Gaussian (or nearly Gaussian) variables of zero mean.
In the Gaussian approximation, the distinguishing statistical properties of the \E{i} are therefore encoded in the allowed values of the \(\vkn\) and in the correlation matrix
\begin{equation}
    C^{\E{i}}_{\vkn \vknp} \equiv \left\langle \tphiF (\vkn) \, \tphiF (\vknp)^* \right\rangle \quad \forall \, \vkn, \vknp \in \reclat(\Ehom{i})\,.
\end{equation}

This correlation matrix is not itself an observable, but the statistical properties of all observables can be expressed in terms of $C^{\E{i}}_{\vkn \vknp}$
and compared for the 18 different Euclidean topologies as functions of the parameters that characterize manifolds of those topologies and observer-locations in them.

\subsection{Observable signatures of topology in three-dimensional surveys}

If the shortest distance around the Universe through us is smaller than the diameter of volumes that we probe with our observations, then we would be able to look directly for voxels (or pixels) that are perfectly correlated (modulo decorrelating line-of-sight effects) because we would be observing the same locations in the Universe in different directions. 
Previous searches for cosmic topology using the ``circles in the sky'' signature for the CMB have established that this shortest distance is greater than $98.5\%$ of the diameter of the LSS \cite{Vaudrevange:2012da}.  
A non-CMB survey would have to probe to at least $z \simeq 500$ in order to directly observe these multiple paths to the same cosmic location.\footnote{
    Observing the same location in two directions requires the survey diameter to exceed the length of the shortest closed geodesic through us, i.e.\ a comoving radius of at least \(98.5\%\) of \(\chi(z_{\textsc{lss}}) = \dlss/2\). 
    Assuming \(h = 0.67, \Omega_{\mathrm{m}} = 0.31\) and \(\Omega_{\Lambda} = 0.69\), such a radius corresponds to a redshift of \(z \simeq 500\) --- see \rcite{Wright:2006up}.
}

Under the assumption that topology-induced correlations arise because of the effect of topology on the eigenmodes of the Laplacian, future searches for topological signatures in cosmological data are probably best conducted in harmonic (Fourier) space rather than configuration (real) space.
In real space, (absent self-intersecting observational volumes) the statistical information is distributed widely across the entries of the voxel-voxel correlation matrix, making it difficult to isolate subtle topological effects. 
In contrast, in Fourier space, wavevector discretization is naturally represented as the localization of the wavevectors onto the reciprocal lattice, and correlations are confined to well-defined Fourier-mode pairs. 

We cannot observe any of the fields $\phi$ over the entire manifold \E{i}, but rather in a restricted subvolume that will eventually depend on each individual survey that observes or estimates the aforementioned field. 
Functionally, this observational constraint amounts to redefining the scalar field by multiplying it by a window function $W(\vx)$ that vanishes outside the observation volume:
\begin{equation}
    \phi_{\obs}(\vx) = \phi(\vx)\cdot W(\vx)\,.
\end{equation}
Then, by the Fourier convolution theorem, the Fourier mode amplitudes of \(\phi_{\obs}(\vx)\) are given by
\begin{equation}
    \tphi_{\obs}(\q) = \frac{1}{V_{\E{i}}}\sum_{\vkn\in \reclat(\Ehom{i})} \tphiF (\vkn)\,\tW(\q - \vkn)\,,
\end{equation}
where \tW\ is the Fourier transform of the window function. 
Hereafter, we drop the \emph{obs} subscript from the field \(\tphi\). 
To avoid confusion, when evaluated at wavevectors \(\vkn\) the field \(\tphi\) refers to the Fourier transform of the globally defined version, whereas when evaluated at wavevector \(\q\) it refers to the harmonic amplitude of the observed field.

We neglect any smoothing on small scales (e.g., due to instrument or survey resolution), which will not have any significant effect on our results since we only consider modes with large enough wavelengths to be in the linear regime.

The statistical properties of \(\tphi(\q)\) are directly linked to those of \(\tphi(\vkn)\).
The expectation value $\langle\tphi(\q)\rangle$ vanishes as long as $\langle\tphi(\vkn)\rangle$ does, whereas their two-point functions are related via
\begin{equation}\label{eqn:cqq_general}
    \bigl\langle\tphi(\q) \, \tphi^{*}(\qp)\bigr\rangle \equiv C^{\E{i}}(\q, \qp) = \frac{1}{V_{\E{i}}^{2}} \sum_{\vkn, \, \vknp \smallin \reclat\left(\Ehom{i}\right)} C^{\E{i}}_{\vkn, \vknp}\tW(\q-\vkn) \tW(\qp-\vknp)^*\,.
\end{equation}
Under the assumption of Gaussian fluctuations, \(C^{\E{i}}_{\vkn, \vknp}\) encodes all the statistical information of the field \(\tphi(\vkn)\). 
Such information will depend both on the underlying cosmology (i.e., on the field content and gravitational theory of the model) and on the topology of the manifold $M$. 
In \rcite{COMPACT:2023rkp}, concise expressions for these quantities were provided for the orientable Euclidean topologies, retaining the standard isotropic power spectrum and highlighting the correlation structure arising from purely global, topological effects.

The quantity \(C^{\E{i}} (\q, \qp)\) might be thought of as a smeared or smoothed version of the pure correlation matrix $C^{\E{i}}_{\vkn, \vknp}$. 
In particular, pairs of modes with vanishing theoretical correlation can become correlated due to the presence of the finite observer window, the strength of such ``spurious'' correlation being modulated by the distance between the modes in Fourier space.~\footnote{This is analogous to the mode-coupling introduced by foreground masks in CMB analysis, whose side effect is to correlate the (otherwise statistically independent) angular scales \(\ell\) and \(\ell^{\prime}\) --- see \rcite{Hivon:2001jp}.} 
This is evident in light of \cref{eqn:cqq_general}. 
The window function \tW\ is generally a decreasing function of the norm of its argument--it must tend to zero at infinity, given the finite nature of the observable volume. 
However, \(\tW\) need not be a \textit{monotonically} decreasing function (see for instance \cref{eqn:Wsinc}), so the coupling between modes does not simply decrease with the distance between their wavevectors. 

This smearing phenomenon is common to both topologically trivial and non-trivial universes; it is, in fact, a basic result in signal processing (e.g., see Ref.~\cite{Harris1978}). 
A key prediction of many non-trivial Euclidean topologies is the presence of correlations between topology modes $\vkn, \vknp$ that are rotations of one another; in other words, the covariance matrix $C^{\E{i}}_{\vkn \vknp}$ contains terms proportional to \(\Kdelta_{\vkn \left(\mat{M}^T \vknp\right)}\), where \(\mat{M} \in O(3)\). 
(See \cref{eqn:E3FourierCovarianceStandardConvention} for such an example.) 
Such correlations are usually modulated by phase factors controlled by the size of the manifolds. 
During the smearing process that occurs when going from $C^{\E{i}}_{\vkn \vknp}$ to $C^{\E{i}}(\q, \qp)$ the phase factors can potentially suppress the off-diagonal correlations, decreasing our ability to detect the topology.
This happens because the larger the manifold's physical size, the faster the phase factors oscillate. 

\subsection{Covariance matrices of some Euclidean manifolds}\label{ssec:covmat_Ei}

In this subsection, we give explicit formulae for the \(\q\)-mode correlation matrices $C^{\E{i}}(\q, \qp)$ for several Euclidean manifolds. We will limit our attention to a selection of representative manifold classes that showcase most of the interesting phenomenology: 
the homogeneous three-torus \E{1}, which exhibits the effect of discretization of eigenmodes common to all compact manifolds;
the quarter-turn space \E{3}, one of the simplest examples with off-diagonal mode correlations; and the usual covering space \E{18}, to which we will compare the previous two cases.

The general lesson is that, when performing the summation in \cref{eqn:cqq_general}, the topological information contained in the $C^{\E{i}}_{\vkn,\vknp}$ matrix can be reorganized to obtain a single (much more simple) sum involving the associated \E{1} eigenmodes  and a mode-mixing kernel $G^{\E{i}}$ that depends only on the manifold details and the window function, but not on the parameters of the background flat FLRW cosmology (a similar separation would be possible in each of the FLRW geometries, with appropriately modified reciprocal lattices).
We will therefore be able to write
\begin{equation}\label{eqn:Cqq_general_simplified}
    C^{\E{i}} (\q, \qp) = \frac{1}{V_{\Ehom{i}}} \sum_{\vkn \smallin \reclat\left(\Ehom{i}\right)} P(\vkn)\, G^{\E{i}}(\vkn, \q, \qp, \vxo)\,.
\end{equation}
Here $P(\vkn)$ denotes the power spectrum of the field $\tphiF (\vkn)$ and contains all the  information about the background Euclidean cosmology. It can generally be split into a product of the dimensionless primordial power spectrum $\PR(\kn)$ and some transfer functions $\Delta(\vkn)$ that encode the evolution of the fields,\footnote{In this paper we have focused on auto-correlations of a single field, but the results presented here can be easily generalized to cross-correlations of different fields by changing $\vert\Delta(\vkn)\vert^{2}$ to $\Delta^X(\vkn)\Delta^Y(\vkn)^*$, where $X,Y$ label the different fields.}
\begin{equation}\label{eqn:ps_relations}
    P(\vkn) = \frac{2\pi^2} {\kn^{3}} \, \PR(\kn) \, \big| \Delta(\vkn) \big|^{2} \,.
\end{equation}

For each topology class we first recall some basic properties, such as the parameters defining a specific manifold of that topology, the allowed wavevector structure and the associated $C^{\E{i}}_{\vkn,\vknp}$ matrix. The reader is referred to \rcite{COMPACT:2023rkp} for a more thorough description of each topology class.

\subsubsection{The three-torus \E{1}}
Any three-torus is fully determined by three linearly independent translation vectors, whose lengths we name $L_x,L_y$ and $L_z$. 
For the illustrative purposes of this paper, we adopt the simplifying assumption that these vectors are orthogonal to each other. 
Dropping this assumption makes the numerics somewhat more involved, but it makes no great conceptual difference.

It is a well-known result that the allowed wavevectors \(\vkn\) in such a three-torus lie on the reciprocal lattice induced by the translation vectors and can therefore be labeled by triplets of integers \(\vn = (n_x, n_y, n_z)\):
\begin{equation}
\vkn = 2 \pi \left(\frac{\nx}{L_x}, \frac{\ny}{L_y}, \frac{\nz}{L_z}\right)\,.
\end{equation}
The correlation matrix of mode amplitudes contains only diagonal non-zero elements:
\begin{empheq}{align}\label{eqn:CE1XY}
    C^{\E{1}}_{\vkn \vknp} = V_{\E{1}} P(\vkn) \, \Kdelta_{\vkn\vknp} \,.
\end{empheq}
Note that this correlation is defined only for wavevectors on the reciprocal lattice, \(\vkn, \vknp \in \reclat(\E{1})\); for any pair where at least one wavevector lies off the lattice the correlation vanishes identically, since only lattice modes are physical degrees of freedom on this topology. 
This is the biggest contrast with the covering-space case below, where the analogous \(\delta^{(D)}(\vk - \vkp)\) allows the full continuum of \(\vk\). 
Substituting \cref{eqn:CE1XY} into \cref{eqn:cqq_general} and making use of the fact that every \E{1} topology is trivially its own associated \E{1} (i.e., $\E{1} = \E{1}^{(1)}$), the observed mode-mode correlation is:
\begin{empheq}[box=\fbox]{align}\label{eqn:Cqq_E1}
    C^{\E{1}} (\q, \qp) = \frac{1}{V_{E_1}} \sum_{\vkn} P(\vkn) \, \tW(\q-\vkn) \, \tW(\qp-\vkn)^*\,.
\end{empheq}
\subsubsection{The quarter-turn space \E{3}}\label{subsec:quarter-turn-space}
The quarter-turn space \E{3} is characterized by two length scales, \(\L{A}\) and \(\L{z}\). 
Because this topology is statistically inhomogeneous, the covariance matrix also depends on the observer's position relative to the corkscrew axis. 
Without loss of generality, we choose coordinates in which this axis is parallel to the \(\unitvec{z}\)-axis. 
The generators consist of two orthogonal translations in the transverse plane, 
\begin{equation}
    \T{A_{1}}{\E{3}} = \L{A} \, (1, 0, 0)^{T},
    \qquad
    \T{A_{2}}{\E{3}} = \L{A} \, (0, 1, 0)^{T},
\end{equation}
and a corkscrew motion
\begin{equation}
    \g{B}{\E{3}} : \quad \vx \to
    \vxo +
    \mat{M}^{\E{3}}_B(\vx - \vxo)
    + \L{z} \, (0, 0, 1)^{T} \,,
    \qquad
    \mat{M}^{\E{3}}_B=
    \begin{pmatrix}
        0 & 1 & 0\\
       -1 & 0 & 0\\
        0 & 0 & 1
    \end{pmatrix},
\end{equation}
where \(\M{B}{\E{3}}\) represents a rotation by \(-\pi/2\) about the \(\unitvec{z}\)-axis. 
With the observer placed at the coordinate origin, the vector \(\vxo = (x_{0}, y_{0}, 0)\) gives the transverse position of the corkscrew relative to the observer.
We note that the corkscrew motion \(\g{B}{\E{3}}\) is independent from the \(z\)-component of the observer position \(z_{0}\). 
As anticipated in \autoref{ssec:top_fundamentals}, the Fourier mode covariance matrix \(C^{\E{3}}_{\vkn \vknp}\) is not diagonal for the quarter-turn space \E{3}:
\begin{empheq}{align}
\begin{aligned}\label{eqn:E3FourierCovarianceStandardConvention}
    C^{\E{3}}_{\vkn\vknp} = {} & V_{\E{3}} \, P(\vkn) \, e^{\I (\vknp - \vkn) \cdot \vxo}
        \left[
            \sum_{\vec{\tilde{n}} \smallin \setN^{\E{3}}_1}
                 \Kdelta_{\vkn \vktn}
                 \Kdelta_{\vknp \vktn}
        \right. \\
    & \left. \quad\quad
            {} + \frac{1}{4} \sum_{\vtn \smallin \setN^{\E{3}}_4}
                \sum_{a = 0}^{3} \sum_{b = 0}^{3} 
                e^{\I \vktn \cdot (\vec{T}^{(a)} - \vec{T}^{(b)})}
                 \Kdelta_{\vkn\left(\left[\left(\M{B}{\E{3}}\right)^{T}\right]^{a\vphantom{b}} \vktn \right)}
                 \Kdelta_{\vknp\left(\left[\left(\M{B}{\E{3}}\right)^{T}\right]^{b} \vktn \right)} 
         \right] \,.
\end{aligned}
\end{empheq}
%

The topology induces perfect 
correlations between a wavevector \(\vkn\) and its rotated counterparts $\iMT \vkn$, $\iMTp{2} \vkn$, and $\iMTp{3} \vkn$;\footnote{
    Here and below we drop the subscript $B$ and the superscript \E{3} on $\mat{M}$ where the reader will surely infer their presence.}
 this is encoded in the Kronecker deltas in the second line of the previous equation. 
Moreover, the covariance matrix depends on where the observer is located with respect to the axis of the corkscrew motion, here manifested via the dependence on $\vxo$. 
The index sets $\setN^{\E{3}}_1$ and $\setN^{\E{3}}_4$ contain the labels of the allowed eigenmodes. For further detail we refer the reader to \cite{COMPACT:2023rkp}. 
The observed mode-mode correlation matrix is given by
\begin{empheq}[box=\fbox]{align}\label{eqn:Cqq_E3}
    C^{\E{3}} (\q, \qp) = \frac{1}{V_{\E{3}^{(1)}}} \sum_{\vn} P(\kn) G^{\E{3}} (\vkn, \q, \qp, \vxo) \,,
\end{empheq}
where 
\begin{align}\label{eqn:GE3}
\begin{aligned}
    G^{\E{3}} (\vkn, \q, \qp, \vxo) & = \tW(\q - \vkn) \, \tWconj\!\left(\qp - \vkn\right) \\
        & \quad\quad + \I^{-\nz} \, e^{{\I \vkn \cdot}[(\iM - \identity)\vxo]} \, \tW(\q - \vkn) \tWconj(\qp - \iMT \vkn) \\
        & \quad\quad\quad\quad + (-1)^{\nz} e^{\I \vkn \cdot[(\iM^2 - \identity)\vxo]} \, \tW(\q - \vkn) \, \tWconj(\qp - \iMTp{2} \vkn) \\
        & \quad\quad\quad\quad\quad\quad + \I^{\nz}\,e^{\I \vkn \cdot [(\iM^3 - \identity)\vxo]}\,\tW(\q - \vkn) \, \tWconj(\qp - \iMTp{3}\vkn) \,.
\end{aligned}        
\end{align}
The first term is the standard diagonal contribution found in \E{1}. 
The subsequent terms represent the off-diagonal couplings induced by topology: a mode \(\q\) is correlated with $\qp$ if the latter is ``close'' to a rotation of \(\q\) by a multiple of $\pi/2$ about the \(z\) axis (because $\M{B}{\E{3}}$ is a rotation by $-\pi/2$ about the \(z\) axis).
The phase factors depend on the relative position of the observer with respect to the axis of rotation through the vector $\vxo$ (which points from the former to the latter), thus breaking statistical homogeneity. 
This demonstrates how the topological signal is modulated by the observer's location within the fundamental domain.

Upon inspection of the kernel \(G^{\E{3}}\), it could seem that off-diagonal correlations between modes coupled by topology could be as large as entries in the diagonal, since each of the four summands in \cref{eqn:GE3} is of the same order of magnitude.
However, when performing the sum over $\vn$, if the dimensions of the survey are small compared to those of the manifold, the rapidly oscillating phase factors sitting in front of the last three summands smear out the contribution to the off-diagonal terms. 
The strength of that smearing is thus modulated by the relative size of the survey volume compared to the size of the \E{3} manifold.

\subsubsection{The covering space \E{18}}
Because \(\E{18}\) is globally homogeneous and isotropic, the covariance matrix does not depend on \(\vxo\). The allowed wavenumbers form a continuum:
\begin{empheq}{align}\label{eqn:E18FourierCovarianceStandardConvention}
    C^{\E{18}}_{\vk \vkp} = (2\pi)^3 P(k) \, \Ddelta(\vk - \vkp)\,.
\end{empheq}
The observed mode-mode correlation in this case is given by
\begin{empheq}[box=\fbox]{align}\label{eqn:CE18}
    C^{\E{18}}(\q, \qp) = \frac{1}{(2\pi)^{3}} \int\!\ddc \vec{k} \, P(k) \, \tW(\q - \vk) \, \tW^{*}(\qp - \vk)\,.
\end{empheq}

If the chosen observer window lacks spherical symmetry (which will be the case in \cref{sec:methodology}), the integrals appearing in \cref{eqn:CE18} do not factorize into simpler one-dimensional forms. In contrast to spherically symmetric windows, where the angular dependence can be integrated analytically, a window function with generic geometry couples the three spatial directions, requiring a full three-dimensional numerical evaluation of the covariance. 
In such cases, for computational practicality, we will approximate \cref{eqn:CE18} by the covariance matrix of a sufficiently large cubic \E{1} topology. 
This approximation is well justified in practice: most simulated realizations of the covering space are such \E{1} realizations, and what makes them behave as \E{18} is simply that the \E{1} cube is chosen to be large enough that the lattice spacing $2\pi/L$ is much finer than any scale of interest (e.g.~\cite{Hockney:1988,Rasera:2013xfa,Schneider:2015yka}). 
The discreteness of the spectrum then becomes numerically irrelevant. Concretely, we require \(L\) to be much larger than the largest wavelength probed by the survey, i.e. \(L\gg 2\pi/k\), so that the lattice sum in \cref{eqn:CE1XY} approximates the continuum integral in \cref{eqn:CE18} to the desired precision.

\section{Methodology}
\label{sec:methodology}
In this Section we present specific choices for various quantities and functions described in \cref{sec:theory}. Given the novelty of the work presented here, simple choices have been made for a majority of them, and more realistic settings and further extensions that will be required for future forecasts will be studied in an upcoming paper 

\subsection{The observer window function}
\label{ssec:obswin}

The window function is the only ingredient in our formalism that depends on the particular survey rather than on the underlying cosmological model or topology. 
Since our aim in this work is to isolate and understand the topological signatures themselves, we deliberately adopt a simple observer window for which the mode-coupling can be more easily studied.
Specifically, we choose $W(\vx)$ to be a cubic top-hat of side length $\Lobs$ centered on the observer:
\begin{equation}
W(\vx) = \begin{cases}
    1 & \text{ if }-\Lobs/2\leq x,y,z \leq \Lobs/2\,, \\
    0 & \text{ otherwise. }
\end{cases}
\label{tophat}
\end{equation}
Furthermore, for computational convenience we assume that the cube is aligned with the coordinate axes used to describe the manifold.\footnote{The inclusion of a cube with arbitrary alignment requires three additional parameters (i.e., three Euler angles) to specify the overall orientation of the fundamental domain relative to, say, the survey frame. This adds to the already existing list of topological parameters, since a complete detection of topology would required knowledge of not only its length scales and shape, but also of its overall orientation relative to the geometry of a realistic survey. The extent to which the inclusion of these parameters impact detectability is left for future study.} 
With these choices, the Fourier transform of the window factorizes into products of one-dimensional functions, 
\begin{equation}\label{eqn:Wsinc}
    \tW(\q) = \Lobs^{3} \sinc(q_x\Lobs/2) \sinc(q_y\Lobs/2) \sinc(q_z\Lobs/2) \,.
\end{equation}
The separability of Eq.~(\ref{eqn:Wsinc}) is the principal advantage of the cubic window.
Every mode-coupling kernel appearing in the covariance matrices now factorizes into products of one-dimensional sinc functions, allowing efficient numerical evaluation even for the off-diagonal covariance matrix entries.

Real galaxy surveys, however, are more naturally described by approximately spherical or light-cone geometries, together with angular masks and non-uniform selection functions. 
Such windows are not separable in Cartesian coordinates (which are naturally adapted to the symmetries of the topologies) and therefore require substantially more expensive numerical calculations. 
Nevertheless, the formalism developed in Section~\ref{sec:theory} is completely general and applies to arbitrary window functions through their Fourier transform \(\tW(\q)\). 
The cubic window should therefore be seen as a convenient first benchmark that isolates the impact of topology while retaining the essential physics of finite-volume mode coupling. 
More realistic survey geometries will be investigated in future work.

Thus, the only relevant scales of this problem are the size of the topology and the size of the observer window function. For a scale-invariant power spectrum (see \Cref{ssec:ps}), only the ratio between these scales will matter for detection (i.e., how much of the topology size is inside the observer window). In the next section, we will study how to compare the constraints with the ones obtained from the CMB, which introduces another scale to the problem, $\LLSS$.

\subsection{Comparison to the CMB}\label{ssec:comp_CMB}
In order to compare the amount of information about cosmic topology that can be extracted from the matter distribution versus the CMB, we parametrize the size of the observed window function through the dimensionless ratio
\begin{equation}\label{eqn:fobs_def}
    \fobs \equiv \frac{\Lobs}{\LLSS} \,.
\end{equation}
The value of $\fobs$ depends on how one chooses to relate the idealized cubic survey to the spherical volume enclosed by the LSS.
Since there is no unique prescription, we consider three representative choices throughout this work,
\begin{equation}\label{eqn:fobs_values}
\fobs = \begin{cases}
    1, & \text{cube that circumscribes LSS},\\[2mm]
     0.935, & \text{\E{1} information-matched benchmark},\\[2mm]
    \left(\dfrac{\pi}{6}\right)^{1/3} \approx 0.806, & \text{cube with same volume as interior of LSS}.
\end{cases}
\end{equation}
The first and third choices have simple geometric interpretations. 
The case \(\fobs = 1\) corresponds to a cube whose side length equals the diameter of the LSS, so that the spherical last-scattering volume is inscribed within the observer box. 
The choice \(\fobs = (\pi/6)^{\nicefrac{1}{3}} \simeq 0.806\) instead makes the cube and the LSS enclose the same volume. 
The intermediate value, \(\fobs \simeq 0.935\), is motivated by information content rather than geometry: since a three-dimensional survey probes the entire observable volume, whereas the CMB samples mostly its two-dimensional boundary, one expects the former to contain at least as much information about spatial topology as the latter. 
We therefore determine $\fobs$ iteratively so that, for the simplest topology (\E{1}), the Kullback--Leibler divergence (see \Cref{ssec:kl}) of the three-dimensional approach matches that of the CMB at a topology scale $L \simeq 1.01\,\LLSS$.\footnote{
    So that we can most readily compare results for \E{3} to those for \E{1}, we choose not to change the size of the information-matched cube when we study the \E{3} topology.
}
This provides a useful benchmark that provides a lower limit on the amount of information available inside the LSS, separating conservative and optimistic assumptions about the effective survey volume.

We stress that the proposed scenarios are not realistic, observationally speaking, given the impossibility for a survey to actually observe such idealized geometries that pierce through the LSS. 
However, these selections allow us to target specific questions concerning the amount of information available on the spatial topology of the Universe in three-dimensional data compared to the CMB. 
The choice \(\fobs = 1\) represents an optimistic scenario, in which the observer box retains all Fourier modes with wavelengths up to the diameter of the LSS. 
The equivolume case, \(\fobs = (\pi/6)^{\nicefrac{1}{3}}\), is a conservative benchmark that tests the extent to which the information content is primarily determined by the observed volume. 
Finally, the intermediate value $\fobs\simeq 0.935$ is an empirically calibrated compromise, chosen so that the information content of the simplest three-dimensional topology is comparable to that of the CMB. 
Together, these three cases allow us to assess how sensitive our conclusions are to the assumed effective survey volume without committing to a specific observational strategy.

\subsection{Matched circles and repeated regions}\label{sssec:repeated_regions}

In a multiply connected universe, the surface of last scattering may intersect one or more of its topological images. Since the intersection of two spheres is a circle, an observer would then see pairs of circles on the CMB sky that correspond to the same physical points on the last scattering surface \cite{cornish1998circles}. The temperature and polarization fluctuations along the two matched circles are therefore expected to be correlated, after accounting for the relative phase and orientation induced by the corresponding topological identification.

Matched circles provide a particularly direct signature of cosmic topology because they identify repeated observations of the same physical region, rather than relying only on topology-induced statistical correlations. Their presence depends both on the topology and on the observer's position. The absence of detectable matched circles places lower bounds on the relevant topological length scales \cite{Vaudrevange:2012da}. However, it does not exclude topologies whose fundamental domains are too large, or whose geometry and observer position do not produce such intersections within the observable volume \cite{Petersen:2023a}.

In large-scale-structure surveys, we could observe an analogous effect, where two apparently distant regions of the survey actually correspond to the same physical region, observed from two different directions: we would see a \emph{repeated region}. In these cases, the amount of information about the cosmic topology is significantly higher. In order to compare matter surveys and the CMB fairly, we always choose cases where there are neither matched circles nor repeated regions.

Like in the CMB, we can study what conditions the topology (and the observer location) must satisfy in order to show these repeated regions in an observation. In \E{1}, this condition simplifies to $L < \Lobs$; this means that cubic \E{1} topologies that are smaller than the observation window will show these repeated regions. In \E{3}, the computation is more involved as it also depends on the location of the observer relative to the axis of rotation. Consequently, it is possible to have manifolds with one topology scale, say, \(L_z\), satisfying $L_z<\Lobs$, while still observing no repeated regions for suitable observer positions. Such configurations are especially interesting from the perspective of topology detection: although they evade direct constraints such as repeated regions or matched circles, the small topology scale still induces a strong modification of the allowed Fourier mode structure. They therefore provide an opportunity to study the statistical signatures of topology in the absence of repeated regions. Throughout this work we always choose locations and sizes that ensure no repeated regions are present in the observer window.

\subsection{The observed modes}\label{ssec:obs_modes}

The covariance matrices derived in the previous section are defined for arbitrary Fourier wavevectors \(\q\) and $\qp$. In practice, however, we evaluate them only on the Fourier basis associated with the cubic observer box. The observed modes are therefore
\begin{equation}
\qm = \frac{2\pi}{\Lobs} \, (m_x, m_y, m_z), \qquad m_i\in\mathbb{Z}.
\end{equation}
Rather than considering the full infinite lattice, we retain only the subset satisfying
\begin{equation}
    |m_i| \leq \mmax, \qquad i = \{x,y,z\} \,,
\label{mmax_def}
\end{equation}
where $\mmax$ is chosen as a numerical free parameter of the analysis. For all values of $\mmax$ considered in this paper, the corresponding wavevectors all lie comfortably within the linear regime of structure formation, so that the assumption of Gaussian fluctuations remains well justified.

In order to provide some numerical sense of the typical length scale of these modes, we note that the observed modes satisfy
\begin{equation}
    \lambda_{m} = \dfrac{2\pi}{|\qm|} \gtrsim 1.6 \left(\frac{\fobs}{\mmax/10}\right) \unit{\giga\parsec} \,.
\end{equation}
A typical modern deep galaxy survey with depth $z=2$ will have access to $|\vec{m}|\gtrsim3$.
A hypothetical future survey out to $z\gtrsim5$ would have access to $|\vec{m}|\geq2$ modes, while one reaching $z\gtrsim25$, such as may be possible with line intensity mapping, would have access to $|\vec{m}|\gtrsim\sqrt{2}$.
We find that imposing a minimum value of $|\vec{m}|\geq2$  on the observer modes typically leads to a $\sim20-25\%$ reduction in the KL.
Ongoing galaxy surveys will therefore have access to many of the modes that contain significant topological information and may, especially in cross-correlation and in combination with the CMB or other high-redshift probes like future all-sky line intensity mapping surveys, be valuable for discovering or constraining cosmic topology. 

For reasons discussed in \autoref{sec:information} and in \cref{app:integralconstraint}, we exclude the $\q=\vz$ mode from the our set of observed modes.
Finally, because the underlying field is real, the Fourier amplitudes satisfy
\begin{equation}
    \tphi(-\q) = \tphi(\q)^{*}.
\end{equation}
The covariance matrices therefore contain redundant information if both \(\q\) and $-\q$ are included.
Throughout this paper we nevertheless retain the complete set of wavevectors, as this simplifies the notation and the full covariance matrix is the natural object to build the relevant probability distribution function.

\subsection{Power spectrum}
\label{ssec:ps}
As argued in \Cref{sec:information}, we focus on perturbations on a primordial spatial hypersurface. Therefore, we take as the dimensionful power spectrum \(P(\vkn)\) to be plugged into \cref{eqn:Cqq_general_simplified} the usual nearly scale invariant Harrison-Zeldovich spectrum,
\begin{equation}
    P_{\textsc{hs}}(\vkn) = \frac{2\pi^{2}}{\kn^{3}} \, \PR_{\rm HS}(\kn) = \frac{2\pi^2 \As}{\kn^{3}}\left(\frac{\kn}{\ks}\right)^{\!\!\ns-1}=\left(\frac{2\pi^2 \As}{\ks^{\ns-1}}\right)\,\kn^{\ns-4}\,.
\end{equation}
Compared to \cref{eqn:ps_relations}, this choice is equivalent to setting the transfer functions $|\Delta(\vkn)|^{2}$ to unity and assuming the typical slow roll inflationary prediction for the power spectrum of the primordial curvature perturbation $\mathcal{R}$. 
It should be emphasized that the absence of a modification to the primordial curvature power spectrum is an assumption rather than an established result, as the details of the impact of nontrivial topology on the primordial spectrum has not yet been established.
For an assessment of the impact on detectability that modifying the primordial power spectrum in a non-trivial topology could have, see \rcite{COMPACT:2026pqs}.

\subsection{Computation of covariance matrices}
Despite their apparent simplicity, numerical evaluation of the covariance matrices \(C^{\E{i}}(\q, \qp)\) requires special attention. 
In this Section we summarize the choices we made in our implementation of the sums over the reciprocal lattice. 

The covariance matrices in Eqs.~\cref{eqn:Cqq_E1} and~\cref{eqn:Cqq_E3} involve sums over the reciprocal lattice of the corresponding topology. 
These lattices are infinite, but the sums may be truncated safely, since the window function strongly suppresses the contribution of topology modes \(\vkn\) lying far from the observed wavevectors \(\q\) and \(\qp\).\footnote{The primordial power spectrum contributes to this suppression as well, but for sufficiently regular spectra the window function is the dominant effect.}

We therefore truncate the lattice at a maximum integer threshold \(\nmax\), corresponding to a maximum topology box wavenumber \(\kmax\). 
In practice, we adopt the pragmatic approach of choosing \(\nmax\) such that the quantity of interest has converged numerically and is demonstrably insensitive to any further increase. 
This criterion is deliberately weaker than demanding convergence of every individual entry of the covariance matrix: the Kullback--Leibler divergence presented in \autoref{ssec:kl}, for instance, stabilizes at values of \(\nmax\) far below those required for entry-wise convergence. 

In our implementation of this problem, we compute the \(\sinc\)-factors in these expressions over the (distinct) \(\vkn\) modes, which are then fed into a computation of the (independent) \(\qm, \qmp\) pairs. 
The space of each of these is about an order of magnitude smaller than their dense representation, which gives us sufficient computational advantage to make the problem tractable.

\section{Results}\label{sec:results}

In this section we quantify the amount of information about cosmic topology contained in ideal three-dimensional observations, namely of the CPS discussed in \autoref{sec:information}, and compare it with that available from the CMB. 
The information content is measured using the Kullback--Leibler divergence between a survey done in a topologically non-trivial manifold and the same survey done in the covering space.
We begin by briefly reviewing the KL divergence and its interpretation. 
We then represent results for the homogeneous cubic three-torus \E{1} for the three different assumptions on effective survey volume presented in \Cref{ssec:comp_CMB}. 
Finally, we extend the analysis to the statistically inhomogeneous quarter-turn space \E{3}, showing how observer-location dependence and topology-induced correlations between Fourier modes affect the discriminatory power of surveys.

\subsection{KL divergence}\label{ssec:kl}

To quantify the distinguishability of two candidate topologies we employ the KL divergence \cite{kullback1951, kullback1959information}. Given two probability distributions $p$ and $q$ for a random variable $\vx$, it is defined as
\begin{equation}\label{eqn:KL_general_def}
    \DKL (p || q) = \int \dd \vx \,\, p(\vx) \ln \left[\frac{p(\vx)}{q(\vx)} \right]=\mathbb{E}_p\left[\ln\frac{p(\vx)}{q(\vx)}\right] .
\end{equation}
In a cosmological context, the variable $\vx$ can, for instance, be the set of harmonic coefficients $\{a_{\ell m}\}$ describing the CMB sky or the Fourier coefficients $\phi(\vk)$ of the observed matter density field. The KL divergence measures the expected logarithmic likelihood ratio in favor of the distribution $p$ over the distribution $q$ when the data is actually drawn from $p$. Equivalently, it quantifies the amount of information lost when $q$ is used to approximate the true distribution $p$. Thus, swapping the two distributions changes which hypothesis is used to generate the data and therefore the interpretation of the divergence. The KL divergence is always non-negative and vanishes if and only if the two distributions are identical almost everywhere.
For zero-mean Gaussian distributions with covariance matrices $\mathbf C_p$ and $\mathbf C_q$, \cref{eqn:KL_general_def} takes the simplified form
\begin{equation}
    \DKL (p || q) =\frac{1}{2} \sum_{i} (\lambda_{i} - \log\lambda_{i} - 1)\,, \quad \lambda_{i} \in \Spec (\mathbf C_{p} \mathbf C_{q}^{-1})\,,
\end{equation}
where the $\lambda_{i}$ are the eigenvalues of $\mathbf C_p\mathbf C_q^{-1}$. 
Since the covariance matrices completely characterize Gaussian random fields of zero mean,\footnote{This is true for both real Gaussian random fields and complex Gaussian random fields that obey a reality condition.} the KL divergence can be evaluated directly from the covariance matrices derived in the previous sections. Within a frequentist framework, a larger KL divergence generally implies a higher probability of detecting the alternative hypothesis (a non-trivial topology) against the null hypothesis (the covering space).

Throughout this work, we compute  $\DKL (\E{i} || \E{18})$, where \E{i} is one of the non-trivial Euclidean topologies presented in \Cref{ssec:covmat_Ei}. 
This choice addresses the following question: assuming the Universe has non-trivial topology \E{i} (with some specific set of parameters), how much statistical evidence would a typical observation provide against the usual hypothesis of a simply connected universe? Although the reversed quantity \(\DKL (\E{18} || \E{i})\)  could offer some complementary information, we retain the former as our figure of merit.

Finally, we ask at what value of the KL divergence there is sufficient information to distinguish between the two models considered here, namely, trivial and non-trivial topologies. There is no universal threshold for distinguishability, as the appropriate criterion depends on conventional choices, analogous to the familiar $3\sigma$ or $5\sigma$ thresholds for Gaussian statistics. 
A forward KL divergence of order unity is sometimes quoted as a rule of thumb for moderate-to-strong information separation. However, this should be regarded only as a loose guideline (here in units of nats) rather than as an operational detection threshold. In general, a larger KL divergence corresponds to a higher probability of detection. Nevertheless, exceeding unity does not, by itself, indicate that detection becomes practically feasible \cite{Samandar:2026a}.

In this context, it is important to keep in mind that the models can either be defined by fixing their parameters to specific values, corresponding to Dirac delta priors in the Bayesian framework, or by marginalizing over the parameter priors, in which case the marginal likelihoods define the model distributions. Since we currently lack informative priors for these parameters, a more realistic Bayesian analysis that marginalizes over them will generally have lower discriminating power.

\subsection{Three-torus \E{1}}

\begin{figure}[t]
    \centering
    \includegraphics[scale=1]{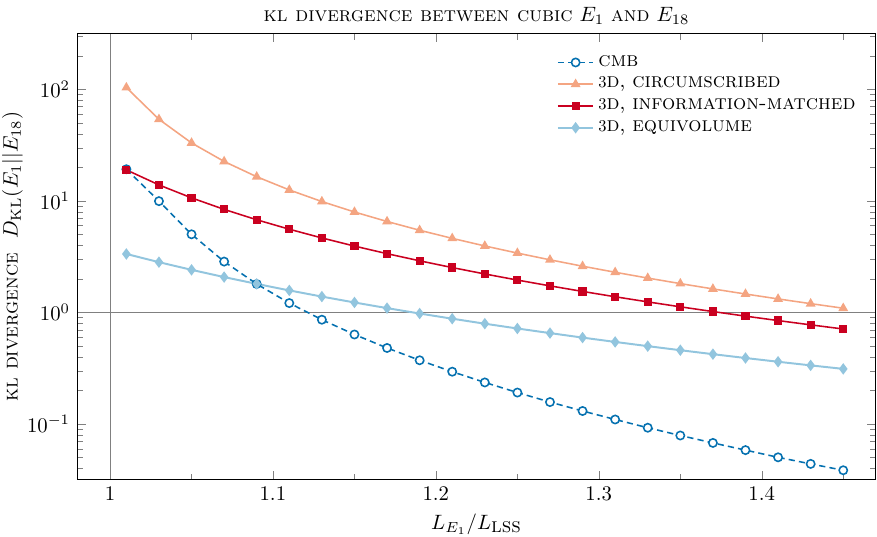}
    \caption{%
        KL divergence of a cubic \E{1} universe against the infinite covering space \E{18}, as a function of the comoving \E{1} sidelength \(L_{\E{1}}\).
        The blue dashed line corresponds to CMB information content, while the three solid lines correspond to three-dimensional observables. Each of the solid lines corresponds to a different survey volume assumption.}\label{fig:E1_plot}
\end{figure}

We begin by evaluating the information content for the simplest Euclidean multiply connected space, the cubic torus \E{1}. \Cref{fig:E1_plot} shows the KL divergence of a cubic $\E{1}$ universe against the covering space \E{18} as a function of the normalized comoving topology scale \(L_{\E{1}} / \LLSS\), for the three choices of observer window introduced in \Cref{ssec:comp_CMB}. The black solid horizontal line at \(\DKL = 1\) represents the widely adopted threshold of distinguishability between models that has been discussed in \Cref{ssec:kl}. 
The black vertical line at $L_{\E{1}}=\LLSS$ represents the boundary between $\E{1}$ manifolds that exhibit matched CMB circles (and thus repeated regions in the $\fobs=1$ scenario) and the ones that do not. 
As previously discussed, we restrict our analysis to configurations where $L_{\E{1}} \geq \Lobs$ in order to avoid the regime containing direct topological intersections (i.e., repeated regions or matched circles), which would yield a much larger KL divergence due to the large correlations expected in these repeated regions.

The maximum multipole of the CMB is set to $\ell_{\rm max}=30$, as the corresponding KL is known to converge by this angular scale when no matched circles are present \cite{COMPACT:2022gbl, COMPACT:2023rkp, COMPACT:2025adc}. 
For  the three-dimensional analysis, we set $\mmax = 10$ (see discussion below). 
The simply connected covering space $\E{18}$ is approximated by a cubic torus with side length $L_{\E{1}}=8\LLSS$, which is large enough to reproduce the continuum limit. 
To compute the covariance matrices that enter the KL computation, the sums have been trunctated at a maximum topology wavenumber of $\kmax = 8\times 10^{-3}\, \unit{\mega\parsec^{-1}}$. 
We have verified that the KL divergence is insensitive to further increases in both $L_{\E{1}}$ and $\kmax$, confirming that the numerical results presented below have converged with respect to those parameters.

The three-dimensional curves shown in \Cref{fig:E1_plot} span a range of plausible assumptions regarding the effective survey volume. 
The optimistic choice, $\fobs = 1$, assumes that all Fourier modes with wavelengths up to the diameter of the LSS are accessible, and therefore provides an upper bound in the information that can be extracted from an ideal three-dimensional survey. 
Accordingly, it yields the largest KL divergence values across all topology scales, well above the CMB curve. 
At the other extreme, the equal-volume prescription with $\fobs=(\pi/6)^{1/3}$ produces the lowest three-dimensional KL values -- comparable to, or even below, those of the CMB, over part of the parameter range. 
This choice should be regarded as excessively pessimistic, since the full three-dimensional volume within the LSS is expected to contain at least as much information about topology as its two-dimensional boundary. 
The equivolume prescription therefore provides a conservative lower bound on the maximum topological information accesible to an ideal three-dimensional survey. 
The third  curve, the intermediate choice with $\fobs=0.935$, was introduced as an information-matched benchmark, chosen to equal the CMB at a value of $L_{\E{1}}$ just larger than $\LLSS$, and then naturally lying always above the CMB curve but below the  $\fobs = 1$ curve.

All the three-dimensional KL curves decrease significantly more slowly than the CMB curve. 
The CMB curve crosses the $\DKL = 1$ threshold line at $L_{\E{1}} \simeq 1.12 \LLSS$, whereas for the 3D curve with the empirically benchmarked value of $\fobs=0.935$ the same crossing occurs at $L_{\E{1}}\simeq 1.37 \LLSS$. 
This represents a gain of roughly $83\%$ in the volume of the largest $\E{1}$ manifold deemed detectable by the previously mentioned criterium. 
Equivalently, an ideal three-dimensional survey remains sensitive to manifolds whose compactification length is substantially larger than the survey itself. 
In the $\fobs=0.935$ scenario, topological signals remain detectable up to $L_{\E{1}}\simeq 1.37 \LLSS$, corresponding to a length approximately $48\%$ larger than the effective survey size. 
By comparison, the CMB becomes insensitive once the topology scale exceeds the diameter of the LSS by only about $12\%$.
Recall that the KL divergence is the expected \textit{logarithm} of the Bayes factor in favor of the true model. 
Therefore, a factor of ten or twenty increase in $\DKL$ (like the ones between CMB and three-dimensional data observed in \Cref{fig:E1_plot}) should not be interpreted as a modest improvement, but rather as a dramatic increase in the amount of information available to discriminate between the topology models being compared.

It is worth noting that this last conclusion is expected to depend only weakly on the absolute size of the survey window. Since the primordial power spectrum is known to be nearly scale invariant (at least in the range of scales considered here, see \Cref{ssec:ps} for further details), rescaling both the survey size and topology scale by the same factor leaves the relative power distribution nearly unchanged. This statement is supported by the observation that for all three-dimensional KL curves in \Cref{fig:E1_plot} the ratio $L^*/\fobs$, where $L^*$ is the value of $L_{\E{1}}/\LLSS$ at which the KL curve crosses the distinguishability threshold, remains constant at about $1.5$. 

The choice of maximum mode number $\mmax$ used in the 3D calculation requires careful consideration, as evaluating the KL divergences for our  covariance matrices becomes computationally expensive very rapidly.  
Even though symmetries of the manifold can be used to speed up calculations, 
the number of independent entries in each covariance matrix still scales as \(\mmax^{6}\) while the runtime to compute KL divergence scales as \(\mmax^{9}\).

\begin{figure}[b]
    \centering
    \includegraphics[scale=1]{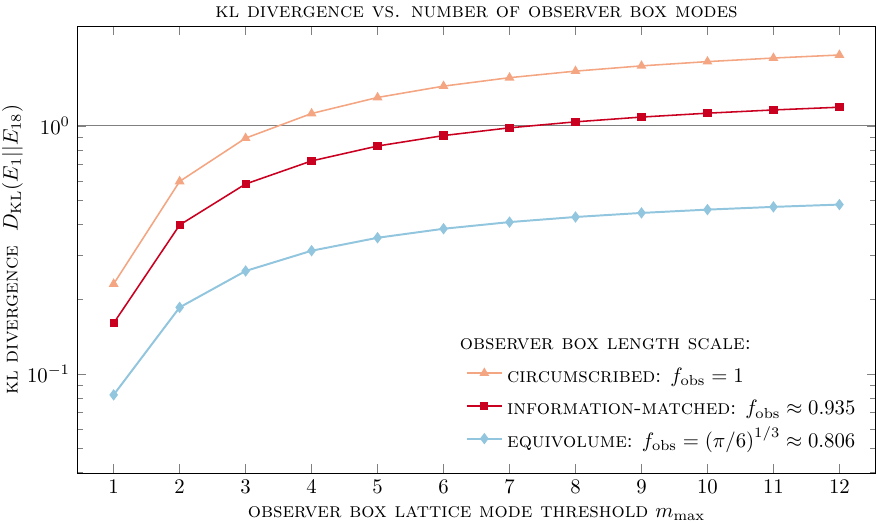}
    \caption{KL divergence of a cubic $\E{1}$ universe with sidelength $L/\LLSS=1.35$ against the infinite covering space $\E{18}$, as a function of the largest included mode index $\mmax$, Eq.~(\ref{mmax_def}). Different lines correspond to different values of $\Lobs/\LLSS$.}\label{fig:Qmax_plot}
\end{figure}

\autoref{fig:Qmax_plot} illustrates how the KL divergence for a fixed $\E{1}$ topology with $L/\LLSS=1.35$ behaves as a function of $\mmax$. 
The discriminatory power rises rapidly for the first few modes, demonstrating that the bulk of the topological signature is encoded in the largest physical scales (i.e., modes with smaller values of $m$). 
As progressively smaller-scale modes are added, the KL divergence continues to increase, but with smaller returns: each additional shell of modes contributes less information than the previous one. Within the range explored here, however, no clear convergence of the KL divergence is observed. 
Thus, the three-dimensional results presented in this work can be regarded as lower limits on the information available in an ideal survey, since including additional (linear) modes can only add to the distinguishability between topologies. 
Our choice of $\mmax$ in Figure \autoref{fig:E1_plot} thus represents a compromise between computational cost and information content.

\subsection{Quarter-turn space $\E{3}$}

The quarter turn space $\E{3}$ provides a simple extension of the cubic-torus analysis. 
While $\E{1}$ isolates the effect of eigenmode discretization associated with compact spatial dimensions, $\E{3}$ exhibits two new related features: statistical inhomogeneity and topology-induced correlation between different Fourier modes. 
As shown in \Cref{ssec:covmat_Ei}, these effects arise from the corkscrew motion defining the manifold. 

\begin{figure}[t]
    \centering
    \includegraphics[scale=1]{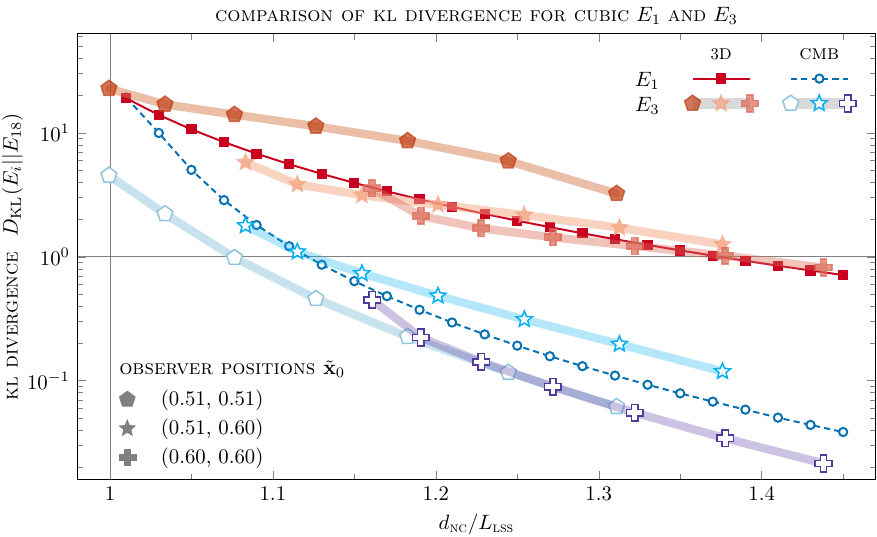}
    \caption{KL divergence of \E{1} and \E{3} universes against the infinite covering space \E{18}, as a function of the ratio of distance to the nearest clone to the diameter of the LSS, $\dnc/\LLSS$, for CMB and 3D observables. For 3D observables, only the $\fobs = 0.935$ scenario is shown. For the \E{3} topology, the scatter is shown for three different observer locations $\tilde{\mathbf{x}}_0$. 
    }
    \label{fig:E3_plot}
\end{figure}

In \Cref{fig:E3_plot}, we extend the analysis and present the KL divergence for both \E{1} and \E{3} topologies as a function of the distance to the nearest clone, $\dnc$.
In cubic \E{1}, $\dnc=L$, while in all the inhomogeneous \E{i}, $\dnc$ varies not just with the parameters of the manifold, but with the location of the observer in the manifold. 
However, $\dnc$ proves to be the best single predictor of the KL divergence. For clarity, we only show the three-dimensional curves corresponding to the information-matched benchmark, \(\fobs \simeq 0.935\), which we adopt as our fiducial three-dimensional scenario throughout this part of the analysis. 
The results have been obtained by fixing the $\E{3}$ translational topology parameter to $L_A=2.5\,\LLSS$ while varying the corkscrew-axis compactification length over the range $0.3\leq L_z/\LLSS\leq 0.9$ in steps of $0.1$.
Choosing $L_A\gg L_z$ isolates the effect of the compact corkscrew direction, whose small size is expected to produce significant modifications to the Fourier structure.

Given that \E{3} is statistically inhomogeneous, the covariance matrix depends on the location of the observer. For the results shown here, we express the transverse observer axis displacement in units of observer box size and define
\begin{equation}
    \txo\equiv\frac{x_0}{\Lobs},
    \qquad
    \tyo\equiv\frac{y_0}{\Lobs}.
\end{equation}
To illustrate this dependence, we plot the KL divergence for three different observer positions $\tilde{\mathbf{x}}_0=(\txo, \tyo) \in \{(0.51, 0.51)$, $(0.60, 0.60)$,  $(0.51, 0.55)\}$ (recall the discussion in \autoref{subsec:quarter-turn-space}).
These choices are designed to allow for values of $L_z$ as small as $0.25\Lobs$ and still avoid repeated regions, provided $L_A$ is big enough ($L_A\geq 2\Lobs$). Such small $L_z$ are of interest because the mode discretization and correlation are both expected to contribute significantly to the KL divergence, as discussed in \Cref{sssec:repeated_regions}.
For \E{3}, the value of $\DKL$ exhibits significant scatter even for the same value of $\dnc$ as the observer position is varied.
For these points, secondary factors such as the number of nearest clones and the distance to the second-nearest clones modulate the value of $\DKL$.

Despite the variability introduced by the freedom of choosing the observer location, the primary conclusion drawn from the $\E{1}$ analysis holds for $\E{3}$ -- for any given distance to the nearest-clone, the information content obtained from the 3D observables is systematically and substantially higher than that from the CMB. 
This shows that the superior discriminatory power of 3D surveys is a robust feature, driven by the number of accessible modes rather than the specific symmetries of the underlying topology.

\section{Discussion and conclusions}
\label{sec:conclusions}

We have performed a first quantitative investigation of the information content of three-dimensional large-scale modes within our Hubble volume pertaining to non-trivial cosmic topology on scales where the shortest closed loop around the Universe through us is greater than the diameter of the last scattering surface.
The spatial boundary conditions imposed by topology discretize the wavevectors of allowed Fourier modes, thereby breaking the isotropy of the background geometry.
In inhomogeneous topologies they also create specific large correlations among modes with wavevectors that are certain discrete rotations (or parity transformations) of one another. 
This is an additional manifestation of statistical isotropy breaking, and also of statistical inhomogeneity.
Both are phenomena that distinguish observations in the non-trivial topology from observations in the covering space.

Starting with the conventional  hypotheses that the amplitudes   of scalar Laplacian eigenmodes are independent random variables with an isotropic nearly scale-free power spectrum, as suggested by inflation,
we computed statistical predictions for  the gauge-invariant scalar curvature field $\cal{R}$ inside a cubic observer volume at last scattering and (with appropriate transfer functions) on the surface of last scattering.
We computed the KL divergence between the statistical predictions for the cubic three-torus $\E{1}$ and those for the covering space $\E{18}$, and similarly between the predictions for the quarter-turn three-torus $\E{3}$ and $\E{18}$.

This KL divergence on the cubic observer volume characterizes the topological information available, in principle, via their relationship to the matter density field, to cosmological galaxy surveys and line intensity mapping surveys.
The KL divergence on the last scattering surfaces represents the information available in the CMB.
We demonstrated that the 3D density field on scales large enough  to remain in the linear regime contains significant imprints of nontrivial topology, well in excess of that contained in the microwave background.
At parameter values where $\DKL$ for the CMB falls below the nominal detection threshold of $1$, $\DKL\simeq5$ for 3D;  and where $\DKL$ for the CMB reliably falls below the detection threshold at or below a distance to the nearest clone of $\dnc\simeq1.1\LLSS$, it reliably stays above 1 until $\dnc\gtrsim1.37\LLSS$ for the 3D survey.
This implies that 3D surveys, alone or in combination with each other and the CMB, will eventually significantly exceed the CMB alone in discovery potential for cosmic topology.

By comparing \E{1}, which has topological information resulting only from eigenvector discretization, with \E{3}, an example case
with significant extra mode-mode correlations, we showed that the topological information depends not only on the overall size of the topological space compared to the size of the Hubble volume, but also on the specific properties of the topological space and (for the case of \E{3}) on the observer location within a fundamental domain. 
Detailed predictions for topological detectability therefore depend on the specific topology, and the parameters specifying the precise manifold. 
Even considering just these two compact topologies, we identify cases where volumes with linear dimension as large as $1.37$ times the last scattering diameter contain a detectably large amount of topology evidence. 

The results of this analysis are, to be sure, idealized compared to realistic surveys. 
The cubic observation volume considered here needs to be replaced by progressively more realistic surveys -- the spherical last scattering ball, our past light cone out to the redshift of last scattering, spherical annuli with redshift ranges appropriate to realistic galaxy surveys and line-intensity mapping surveys. 
The effects of partial sky coverage should also be incorporated. We have used a simple real-space tophat window function, Eq.~(\ref{tophat}), for convenience, but more sophisticated choices could  enhance particular mode correlations in the observed volume and make a given topology more  detectable, at the expense of greater computational complexity. The signal processing literature provides a wealth of window functions and design
strategies to explore \cite{Harris1978}.
Perhaps most importantly, all of the well-known dynamical and photon-propagation effects  that affect the map from $\cal{R}$ on the initial hypersurface at reheating to observables on our past light cone need to be properly incorporated, for example as in \cite{Bonvin:2011bg} but with even more careful attention to preserving all of the available long-wavelength information.

In principle this is straightforward, but we expect that it will be computationally challenging to determine the resulting mode couplings.
Therefore, we will need to develop computationally efficient algorithms for searching the topology parameter spaces. 
For each Euclidean topology, there are up to 11 parameters relating to the size and shape of the manifold, the observer location, and the observer orientation. 
Our initial forays into machine-learning techniques are so-far inconclusive \cite{COMPACT:2024dqe}.
Substantial parameter spaces are not only computationally expensive to search, but increase the chance of mistaking statistical noise fluctuations for a signal and require higher detection thresholds. 

In practice, can we extract this information from realistic future observations? 
Clearly, it will be advantageous to probe as much of the Hubble volume as possible. 
Deep surveys with large sky coverage are clearly best, but the relative importance of higher redshift versus higher signal to noise remain to be explored. 
The goal is to probe as many long-wavelength modes as possible with well-controlled systematics and selection effects that are uniform and do not induce poorly understood mode couplings.
Accurate characterization of survey completeness across the survey will be an important challenge.

Ongoing or planned surveys may be useful, perhaps especially in cross-correlation with the CMB.
LSST will cover approximately half of the sky to $z\simeq3$, but with only photo-z's; smearing in the radial direction due to only statistically constrained redshifts will reduce the detectability of any topological signal of coupling between modes with any radial wavevector components. 
Spectroscopic redshift surveys will not have this issue but are expensive.
The Euclid Wide survey will have spectroscopic redshifts  out to $z\simeq 2$, but covers only about a third of the sky.
DESI is similar, but will be deeper ($z\lesssim3$) for some source classes.

Ideally surveys that reached $z\gtrsim5$ over all the sky would be best.
For this, intensity mapping seems the most promising approach.
SPHEREx expects to reach $z>6$ (and perhaps even $z\simeq9$) with its intensity mapping survey, but only over a few hundred square degrees -- it also plans to trace out the full 3D structure of the universe with a nearly spectroscopic (i.e. $\sim100$ independent wavebands) galaxy survey over 75\% of the sky (after foreground subtraction), but only out to $z\simeq2$.
Nevertheless, ongoing and planned surveys will have access to many of the most informative modes for  cosmic topology, and a detailed analysis of their discovery reach is both warranted and planned.

The gold standard target is a  $21$cm line intensity mapping survey out to high redshifts.  
This has long been touted as the ultimate probe of modes within the observable universe.  
The most daunting challenge has been separating foregrounds -- galactic and anthropogenic -- from cosmological signature.
Using advanced foreground-cleaning techniques,
the Canadian Hydrogen Intensity Mapping Experiment (CHIME) has successfully achieved first cosmological signal detections \cite{CHIME:2023til}, but only by cross-correlating its 21-cm hydrogen data with external spectroscopic galaxy catalogs.  
The Square Kilometre Array Observatory (SKAO)  expects to have  full-sky independent cosmology-grade data out to $z\simeq3$  by 2040 \cite{Elahi:2026ple}.
It remains to be seen whether future decades meet the promise of billions of linear mode amplitudes; if so, this would be the most powerful probe of cosmic topology. 

In principle, the full suite of large-wavelength modes of the matter density field and other tracers of the scalar curvature field, available to us in the volume encompassed by our observable universe is sufficient to extend our discovery potential for cosmic topology well beyond what we will ever be able to do with the cosmic microwave background.
Putting that principle into practice remains a theoretical, calculational and observational challenge.

\appendix

\section{The integral constraint: excluding information from outside the survey}
\label{app:integralconstraint}
Any finite survey can only measure fluctuations relative to its own mean density, not relative to the actual or expected mean on some larger volume.\footnote{
    Of course, that mean can also be measured, and compared to the means of other fields, with consequences for inferences about the cosmological model.}
This imposes an \emph{integral constraint} (IC): 
the mean of the observed fluctuations of a field is defined to vanish on the survey volume. 
As a result, fluctuations on scales comparable to or larger than the survey are partially absorbed into the estimated mean, leading to an artificial suppression of power at the largest scales \cite{Peebles:1976, Peebles:1980,deMattia:2019vdg}. In configuration space, the IC for a generic field $\delta$ is implemented by defining the \emph{convolved, integral-constrained} (CIC) field
\begin{equation}\label{eqn:cic_def}
    \delta^{\text{cic}}(\vx) = W(\vx)
    \left[
    \delta(\vx) - 
    \frac{\int \ddc \vr\,\delta(\vr) W(\vr)}{\int \ddc \vr\,W(\vr)}
    \right],
\end{equation}
which explicitly enforces $\int \ddc \vx\, \delta^{\text{cic}}(\vx) = 0$.
If $W(\vx)$ is uniform within the survey, we note that $\int \ddc \vx\,W(\vx) = \tW(0) = V_{\obs}$.

Taking the Fourier transform of \cref{eqn:cic_def} yields
\begin{equation}
    \widetilde{\delta}^{\text{cic}}(\q)
    = \widetilde{\delta}^{c}(\q)
    - \frac{\tW(\q)}{\tW(0)}\,\widetilde{\delta}^{c}(\vz),
\end{equation}
where $\widetilde{\delta}^{c}(\q) = (2\pi)^{-3} \!\int \ddc \vk\, \widetilde{\delta}(\vk) \tW(\q-\vk)$ is the window-convolved field.
By construction, $\widetilde{\delta}^{\text{cic}}(\vz) = 0$, i.e. the monopole is removed.

The covariance of the IC-corrected modes follows as
\begin{align}\label{eqn:CCIC}
    C^{\text{cic}}(\q,\qp) &\equiv
    \langle \widetilde{\delta}^{\text{cic}}(\q) \widetilde{\delta}^{\text{cic}}(\qp)^* \rangle \nonumber\\
    &= C^{\text{c}}(\q,\qp)
    + \frac{\tW(\q)\tW(\qp)^*}{|\tW(0)|^2}\, C^{\text{c}}(\vz,\vz)
    - \frac{\tW(\q)}{\tW(0)}\, C^{\text{c}}(\vz,\qp)
    - \frac{\tW(\qp)^*}{\tW(0)}\, C^{\text{c}}(\q,\vz) .
\end{align}
Each term in the expression diverges separately due to the contribution from modes with $k \to 0$ if the infrared behavior of the power spectrum is sufficiently steep. In particular, for cosmological initial conditions where the primordial spectrum scales as $P(k)\propto k^{\ns-4}\simeq k^{-3}$, the infrared divergence makes the subtraction enforced by the IC essential.

However, the full combination in \cref{eqn:CCIC} is finite, and grouping the terms under a single integral gives
\begin{equation}
    C^{\text{cic}}(\q, \qp) = \frac{1}{(2\pi)^3} \int \ddc\vk\, P(k)\, F(\q, \qp, \vk),
\end{equation}
with the kernel
\begin{align}
    F(\q, \qp, \vk) &=
    \tW(\q-\vk)\tW(\qp-\vk)^*
    + \frac{\tW(\q)\tW(\qp)^*}{|\tW(0)|^2} |\tW(\vk)|^2 \nonumber\\
    &\quad
    - \frac{\tW(\q) \, \tW(\qp - \vk)^{*} \tW(\vk)^{*}}{\tW(0)}
    - \frac{\tW(\qp)^*\tW(\q - \vk) \tW(\vk)}{\tW(0)}.
\end{align}
This kernel satisfies $F(\q, \qp, \vk=\vec{0})=0$, showing that the unobservable $\vk=\vec{0}$ (mean-density) mode contributes no variance.

The integral constraint thus modifies the covariance at the largest scales, ensuring that the survey-averaged fluctuation vanishes. In general, it removes the $\q = \vec{0}$ mode and suppresses correlations between modes with wavelengths comparable to the survey size.

In this work we have focused on a cubic top-hat observer window, and only studied the modes $\qm$ of that window. With these choices, it can be seen that $\tW(\qm) = 0$ when $\qm \ne 0$, while $\tW(\vz)=\Lobs^3$. Substituting into the general IC-corrected covariance of \cref{eqn:CCIC} shows that the integral constraint therefore affects only the $\q = \vec{0}$ mode, which is explicitly removed by construction. Consequently, for the cubic top-hat and the considered modes, the integral constraint just imposes that the global mean of the perturbation field vanishes, with no residual impact on other Fourier modes:
\begin{equation}
    C^{\text{cic}}(\qm, \qmp) = 
    \begin{cases}
        C^{\text{c}}(\qm, \qmp), & \qm, \qmp \ne \vec{0}, \\[2mm]
        0, & \text{if any of } \qm, \qmp = \vec{0}.
    \end{cases}
\end{equation} 
This simplification is, however, specific to the top-hat window and selected modes. Under more general choices of observer window, or selecting different observed modes, one would need a full treatment of the integral constraint effects in \cref{eqn:CCIC}.

\section*{Acknowledgements}
D.P.M., S.A.\ and G.D.S.\ thank M. Bruni for discussions about issues related to GR. 
CWRU authors thank Daniela Calvetti and Erkki Somersalo for many discussions about computational approaches. 
S.A.\ thanks Pier-Stefano Corasaniti and Alessandro Renzi for discussions. 
G.D.S.\ also thanks D.~Spergel, N.~Cornish, and J.R.~Bond for many relevant conversations over many years.
D.P.M., G.D.S., C.J.C., and A.K.\ acknowledge support from\ NASA\ ATP grant RES240737. 
D.P.M.\ acknowledges support by the Bulgarian National Science Fund programme ``VIHREN--2024'' project No. KP--06--DV/9/17.12.2024. 
M.M.B. acknowledges support by the Spanish Ministry of Science, Innovation and Universities under the FPU predoctoral grant FPU22/02306 and 
by the Spanish Ministry of Science, Innovation and Universities under the FPU predoctoral grant FPU22/02306. 
J.C.D.\ is supported by the Spanish Research Agency (Agencia Estatal de Investigaci\'on), the Ministerio de Ciencia, Innovaci\'on y Universidades, and the European Social Funds through grant JDC2023-052152-I, as part of the Juan de la Cierva programme. 
G.D.S.\ and A.S. acknowledge past support from DOE grant DESC0009946.
Y.A.\ acknowledges support by the Spanish Research Agency (Agencia Estatal de Investigaci\'on)'s grant RYC2020-030193-I/AEI/10.13039/501100011033, by the European Social Fund (Fondo Social Europeo) through the Ram\'{o}n y Cajal programme within the State Plan for Scientific and Technical Research and Innovation (Plan Estatal de Investigaci\'on Cient\'ifica y T\'ecnica y de Innovaci\'on) 2017-2020, by the Spanish Research Agency through the grant IFT Centro de Excelencia Severo Ochoa No CEX2020-001007-S funded by MCIN/AEI/10.13039/501100011033, by the Spanish National Research Council (CSIC) through the Talent Attraction grant 20225AT025, and by the Spanish Research Agency's Consolidaci\'on Investigadora 2024 grant CNS2024-154430.
F.C.G.\ is supported by Ministerio de Ciencia, Innovaci\'on y Universidades, Spain, through a Beatriz Galindo Junior grant BG23/00061 and by an UCOLIDERA grant from the Universidad de Córdoba.
A.H.J.\ acknowledges support from STFC in the UK\@. 
A.N.\ and C.T. are supported by  Richard S.\ Morrison Fellowships. 
R.G.R.\ is supported by Coordena\c{c}\~ao de Aperfei\c{c}oamento de Pessoal de N\'ivel Superior (CAPES).
T.S.P.\ is supported by Funda\c{c}\~ao Arauc\'aria (NAPI Fen\^omenos Extremos do Universo, grant 347/2024 PD\&I).
A.T.\ is supported by the European Union's Horizon Europe research and innovation 
programme under the Marie Sk\l{}odowska-Curie grant agreement No.\ 101126636. 
G.A. is supported by the Spanish Research Agency’s Consolidaci\'on Investigadora 2024 grant CNS2024-154430.
This work is partially supported by the Spanish Research Agency (Agencia Estatal de Investigaci\'on) through the Grant IFT Centro de Excelencia Severo Ochoa No. CEX2020-001007-S, funded by MCIN/AEI/10.13039/501100011033. 
This work is partially funded by the European Commission - NextGenerationEU, through Momentum CSIC Programme: Develop Your Digital Talent. 
This work made use of the High-Performance Computing Resource in the Core Facility for Advanced Research Computing at Case Western Reserve University and the facilities of the Ohio Supercomputing Center.
Numerical calculations have been performed on the Hydra cluster at IFT. We acknowledge HPC support by Emilio Ambite, staff hired under the Generation D initiative, promoted by Red.es, an organisation attached to the Spanish Ministry for Digital Transformation and the Civil Service, for the attraction and retention of talent through grants and training contracts, financed by the Recovery, Transformation and Resilience Plan through the European Union’s Next Generation funds.
We thank CWRU, ICL, the INFN (Italy), the IFT (Madrid), and the U. of Padua, for their hospitality during collaboration research visits.

\bibliographystyle{utphys}
\bibliography{topology}
\end{document}